# Carbon dioxide and amorphous water ices in the debris disk around β Pictoris

S. K. Betti[1*], C. H. Chen[1,2], C. Xie[2], K. Worthen[2], S. Brittain[3], J. Najita[4], I. Rebollido[5], T. Beck[1], B. T. Bolin[6], D. Hines[1], C. Ingebretsen[2], C. Lisse[7], M. D. Perrin[1], L. Pueyo[1], I. Wong[1], and C. Stark[8]

[1]Space Telescope Science Institute; Baltimore, MD, USA.

[2]Department of Physics and Astronomy, Johns Hopkins University; Baltimore, MD, USA.

[3]Department of Physics and Astronomy, Clemson University; Clemson, SC, USA.

[4]US National Science Foundation National Optical-Infrared Astronomy Research Laboratory (NOIRLab); Tucson, AZ, USA.

[5]Centro de Astrobiología (CAB), Consejo Superior de Investigaciones Científicas (CSIC)-Instituto Nacional de Técnica Aeroespacial (INTA); Madrid, Spain.

[6]Eureka Scientific; Oakland, CA, USA.

[7]Space Exploration Sector, Johns Hopkins University Applied Physics Laboratory; Laurel, MD, USA.

[8]NASA Goddard Space Flight Center; Greenbelt, MD, USA.

*Corresponding author. Email: sbetti@stsci.edu

**The star β Pictoris is orbited by a dusty debris disk, within which planet formation is occurring. The disk extends far from the star such that molecules in the outer disk should freeze. We observed the β Pictoris disk in the near-infrared using the James Webb Space Telescope. The spectrum shows features due to amorphous water ice (<7% crystallinity), and carbon dioxide ice in both the $^{12}CO_2$ and $^{13}CO_2$ isotopologues. We infer that the $CO_2$ ice is intimately mixed with the water ice, at a ratio of 1:10. This implies that the ices formed in cold conditions, likely in the interstellar medium during a phase of amorphous $H_2O$-dominated formation. Frequent collisions within the debris disk are required to explain the observed uniform distribution of icy dust grains.**

The young (~20 Myr) (*1*) planetary system β Pictoris (also catalogued as HD 39060) is located 19.6 parsecs (pc) from Earth (*2*). It consists of a central star, an edge-on debris disk, and three known exoplanets. No ices have been directly observed in the β Pictoris disk. Previous work has set an upper limit on $H_2O$ gas of $< 1.1 \times 10^{18}$ kg at 95% confidence (*3*). A photometric search found no evidence of a water ice absorption feature expected at 3 μm (*4*). However, disk conditions imply its existence The disk contains a large reservoir of CO gas (*5*). In the conditions of the β Pictoris disk, this CO gas is expected to photo-dissociate within 50 to 300 years (*5*,*6*), so the CO gas is thought to have been released recently from solid ice by collisions between volatile-rich comets or planetesimals (*7*). The exoplanet β Pictoris b also orbits within the disk; its properties imply that it formed from ice-rich planetesimals (*8*).

Theoretical models have shown that ultraviolet (UV) radiation from the central star could destroy small grains of pure ice faster than they can be produced in planetesimal collisions (*9*). However, ices on dust grains mixed with other molecular species (known as dirty ices) could survive longer (*10*). The 3 μm water ice feature has been observed in one other debris disk, around the star HD 181327, which was interpreted as due to large crystalline water-ice grains (*11*).

Collisions are thought to be common within the β Pictoris debris disk (*12*). Mid-infrared imaging has shown a curved asymmetric dust feature in the southwest arm of the disk (*4*). This was interpreted as 0.1 to 5 μm carbonaceous dust particles produced by a collision ~200 years ago, ~85 astronomical units (au) from the host star (*4*), a little less than midway through the dust disk. Such collisions could also release ice buried within planetesimals.

Ices produce both absorption and scattering features in infrared spectra. Previous observations of ices in protoplanetary disks (PPDs) (*13, 14, 15*), and the HD 181327 debris disk (*11*) found that the most abundant molecules are $H_2O$, $CO_2$ and CO, which all have prominent features at near-infrared wavelengths. The 3 μm OH stretching vibrational mode of $H_2O$ ice produces a distinctive broad feature. $CO_2$ ice has a narrow asymmetric stretching normal mode absorption feature at 4.27 μm for $^{12}CO_2$ and 4.39 μm for $^{13}CO_2$. The strongest CO ice absorption feature is at 4.67 μm (*16*).

## Near-infrared spectroscopy of β Pictoris

We observed β Pictoris on 7 December 2023 with the Near Infrared Spectrograph (NIRSpec) (*17*) on the James Webb Space Telescope (JWST). NIRSpec was used in integral field unit (IFU) mode with the low-resolution prism, which provides a wavelength dependent spectral resolving power of 30 to 400 (nominal is ~100) over the wavelength range 0.6 to 5.2 μm (*18*). To cover the full spatial extent of the disk and mitigate the effect of saturation by the host star, we positioned four IFU mosaic tiles, two on each side of the star (*19*). The resulting combined field-of-view was 5.6″ × 12.8″, with a gap around the position of the host star; at the distance of β Pictoris (19.6 pc), this corresponds to 109 au × 250 au. We additionally observed a reference star in the same observing mode, to determine the instrumental point spread function (PSF). We used reference-star differential imaging (*11, 19)* to subtract the light from the host star.

To search for ices in the disk, we computed a map of the 3 μm water ice feature (*19*). We found that this feature is approximately uniform across the disk, with little variation (Fig. 1C) We therefore extracted integrated spectra for two representative (*19*) locations in the outer disk, defined by 7.8 × 23.5 au oval apertures (Fig. 1A) located ~100 au from the central star, to the northeast (NE) and southwest (SW). After PSF subtraction, we converted the disk flux to disk reflectance, or the fraction of light scattering off of dust particles compared to the stellar light (*19*); the resulting disk reflectance spectra for each aperture are shown in Fig. 1B.

We identify multiple ice features in these disk reflectance spectra. The spectra of the two regions are almost identical, with similar continua and features, including the ice profile shapes and depths (Fig. 2). The broad dip from 2.7 to 3.2 μm is consistent with the 3 μm water feature, if it is from water ice-rich grains. The spectra also show narrower dips at 4.2 to 4.35 μm from $CO_2$ ice. These are from $^{12}CO_2$ at 4.25 μm and $^{13}CO_2$ at 4.39 μm.

There are also emission lines at ~4.6 μm (Fig. 1B), which we attribute to CO gas. Cold CO gas has previously been observed in β Pictoris, in submillimeter emission (*5*), UV absorption (*20*), and near-infrared absorption (*21*). We do not detect any features attributable to solid CO ice, perhaps due to overlapping CO gas emission lines (*19*).

## Band depths and ice abundances

From the observed ice features, we measure the effective optical depths or feature depth, the logarithmic fraction of light at each wavelength that scatters off ice grains (Fig. 2). In optically thick PPDs, multiple scattering angles and temperature gradients can affect the measured depth of scattered light features (*14*). For β Pictoris, we find that the debris disk is optically thin, so each photon experiences only a single scattering event (*19*). If there are any ice-free regions within the disk the ice-containing regions would need stronger ice absorption to produce the same integrated spectra, implying higher optical depths in those regions. For micron sized grains, the ice features are dominated by scattering effects which can affect the band profile and the optical depth determination; we assume these are negligible. For $H_2O$, $^{12}CO_2$, and $^{13}CO_2$, we find maximum effective optical depths of ~0.14, ~0.06, and 0.02, respectively.

We model the optical depth profiles (Fig. 2) by fitting them with skewed Gaussian functions (*19*). Table 1 lists the effective optical depths, peak positions, and half width at half maximum (HWHM) for each ice feature in the NE and SW spectra. We find that the peak wavelength of the $CO_2$ feature is shifted blueward, relative to laboratory transmission spectra of pure crystalline $CO_2$ ice (*22, 23*). For $^{12}CO_2$, the peak position within a 1σ uncertainty is 4.2558±0.0022 μm in the NE arm and 4.2625±0.0015 μm in the SW arm, compared to the peak position of 4.267 to 4.268 μm (*22*) for pure $^{12}CO_2$ ice. For $^{13}CO_2$, the peak position is 4.3893±0.0011 μm in the NE and 4.3882±0.0021 μm in the SW, compared to 4.380 to 4.381 μm (*22*) for pure $^{13}CO_2$ ice, a difference that corresponds to ~4 $cm^{-1}$ in wavenumber. The FWHM of the $^{13}CO_2$ profiles are also broader (~14 to 22 $cm^{-1}$) compared to pure ice (2.3 $cm^{-1}$). We find that this broadening cannot be explained by Dopper shift, as these icy grains travel with velocities low enough that any shift is negligible (<10% of a NIRSpec IFU spectral channel). Broad and shifted $^{13}CO_2$ profiles are known to be produced by molecular interactions in polar matrices (such as $H_2O$ ice) (*24*).

For each ice feature, we measure the total absorbance by integrating the optical depth profiles. Using laboratory measurements of the line strengths, we then calculate the effective column density of ice encountered by photons that scatter off the disk (*19*). We find effective column densities of ~$3 \times 10^{17}$ $cm^{-2}$, ~$3 \times 10^{16}$ $cm^{-2}$, and ~$4 \times 10^{15}$ $cm^{-2}$ for $H_2O$, $^{12}CO_2$, and $^{13}CO_2$, respectively. These column densities indicate a $CO_2$ to $H_2O$ ice column density ratio of 1:9.3±0.1. This corresponds to an abundance (by number) of ~11% $CO_2$ relative to $H_2O$. This is within the range of 2.5 to 30% in Solar System comets (*25*, *26*), and 7 to 20% in PPDs (*26, 27*). Therefore, the ices in the β Pictoris disk might have formed via the same processes, accumulating in water-rich dense molecular clouds in the early stages of star formation (*28*), as in the Solar System and PPDs.

## Composition of the water ice

The shape of the $H_2O$ band profile is affected by the composition and crystallinity of the ice grains (*29*). The observed 3 μm $H_2O$ ice feature absorption band is smooth and broad, with maximum absorption around 2.97 μm (Fig. 2 A & C). We do not detect the 1.65 μm water ice feature or a sharp peak in water ice absorption indicative of crystalline water ice (CWI), the Fresnel peak, at 3.1 μm (Fig. 1B), which are both indicators of crystalline water ice (*29,30*). The lack of crystalline ice signatures, and the peak absorption position near 3.0 μm, indicates that the water ice is mostly amorphous (*16*). Theoretical models predict that the dominant form of water ice in the inner regions of debris disks is crystalline, not amorphous, because heating would cause it to crystallize (*31*). Evidence for this effect has been observed in the HD 181327 debris disk, which has a strong Fresnel peak indicative of crystalline water ice in the inner disk, but no detected amorphous water ice (*11*). Both crystalline and amorphous water ice have been observed in the Solar System (*32*). If water ice condensed in amorphous form early in the star formation process, to remain amorphous it must have remained below ~100 K (*33, 34*) or undergone processes that disrupt the crystal lattice, such as photolysis from high-energy photons, radiolysis from galactic cosmic rays, or sputtering from energetic stellar wind ions (*35*). In the outer disk, primordial amorphous water ice (AWI) could then be common as a result of these effects (*31*).

Fig. 3 compares the measured water ice optical depth profiles to synthetic profiles of AWI and CWI (*19)*. We find that the CWI model (Fig. 3C) predicts a detectable Fresnel peak, which we do not observe. The AWI model (Fig. 3A) is a closer match to the band profile. We also tested models that are aggregates of other ices intimately mixed with $CO_2$, finding that a mixture of either AWI+$CO_2$ (Fig. 3A) or AWI+CWI+$CO_2$ (Fig. 3B) can also reproduce the observed band profiles. We conclude that AWI and $CO_2$ are probably the major components, though cannot exclude the presence of other molecules within the mixture.

AWI must be the dominant phase, but we cannot determine the amount of CWI. To place an upper limit on the crystallinity fraction, we take the best fitting parameters of the AWI+CWI+$CO_2$ model, then vary the volume fraction of AWI until the modeled Fresnel peak optical depth is >3σ from the observed profile (Fig. S7). We find that the water ice is <4% crystalline by total volume, which corresponds to CWI contributing < 7% of the volume of just the ices (3σ upper limits).

**Characterization of the $CO_2$ profile**

The band profiles of the $^{12}CO_2$ and $^{13}CO_2$ features are affected by the ice's composition (*36*) and thermal history (*28*). In proto-stellar environments, $CO_2$ embedded in different ice matrices produces a multiple- component spectral feature (*36, 37*). Similar multi-component $CO_2$ profiles have been observed in PPDs (*38*). Compared to the $^{12}CO_2$ feature at 4.26 μm, the $^{13}CO_2$ band is less affected by saturation or distortion due to grain size or shape effects (*28*). . We therefore use the $^{13}CO_2$ band to determine the dominant $CO_2$ ice matrix by considering the peak position and width (*19*).

Fig. 4A compares the $^{13}CO_2$ feature to laboratory spectra of $CO_2$ ice mixtures at different ratios, which are written as relative ratios of the mixed components (e.g. an equal 1 to 1 mixture of $CO_2$ to CO ice is written as CO:$CO_2$=1:1). The observed feature peaks at longer wavelengths than pure $CO_2$ or a CO:$CO_2$=1:1 mixture. The observed feature is also broader than any of the laboratory spectra, indicating that there are probably multiple ice components (*36*). We find that the closest match is with a mixture of $CO_2$ and amorphous $H_2O$ in a ratio of 1:10 ($CO_2$:$H_2O$=1:10), which was measured at temperatures of 10 to 50 K (Fig. 4B). Although other mixtures are likely possible, this spectrum matches the observed maximum optical depth at 4.392 μm and is consistent with the $CO_2$ to $H_2O$ abundance ratio of ~11% we calculated from the column density ratios of $^{12}CO_2$ and $H_2O$. The laboratory temperature is also consistent with the ~48 K dust temperature at ~100 au found from radiative equilibrium in the blackbody limit from β Pictoris (*39*), and with the ~21 ± 8 K astrophysical ice temperature predicted for a $H_2O$ dominated $^{13}CO_2$ ice band centered at 4.389 μm (*40*). We therefore suggest that a $CO_2$-$H_2O$ mixture is probably the dominant configuration of the $CO_2$ ice with the $CO_2$ ice located in a $H_2O$-rich environment in the β Pictoris disk.

**Origin of the ices**

In the debris disk phase of star formation, ice grains that formed in the earliest and coldest stages of star formation collide, releasing small dust grains that could potentially be detected in scattered light or thermal imaging. The main source of dust production in the outer regions of debris disks is thought to be collisions between planetesimals or comets (*41*). In β Pictoris, the ices are distributed approximately uniformly throughout the disk (Fig. 1C). If a single large recent collision was responsible, then the ice distribution would have traced orbits that pass close to the collision site, as the observed CO gas does (*5*). The uniform distribution of the $CO_2$ and $H_2O$ ices is more consistent with a larger number of smaller collisions between volatile-rich bodies in the β Pictoris disk. We expect such collisions to release small dirty ice grains, which are subsequently heated and blown out of the disk (*4*), limiting their lifetime. Dirty ice grains have a lower rate of photo-desorption compared to pure ices, so could persist for longer in the β Pictoris disk (*10*).

The crystallinity of water ice grains is theoretically expected to depend on the impact velocity, but not the sizes of the colliding bodies (*42*). In the early debris disk stage, impact velocities are expected to be $< 2$ km $s^{-1}$ in the outer β Pictoris disk, which is insufficient to crystallize water ice (*42*). Therefore, the continuous collisions of the parent bodies likely occur at low impact velocities such that the water ice on the dust grains remains amorphous.

The abundances of $H_2O$ and $CO_2$ ices in the β Pictoris disk are similar to those found in PPDs and Solar System minor bodies. Assuming they all have the same origin, this indicates that ices formed early, probably in the interstellar medium (ISM) during an $H_2O$-dominated phase, where $H_2O$ ice mantles begin to accumulate in dense molecular clouds (*28*), or during the proto-stellar envelope phase, one of the earliest stages of star formation, where material from the ISM is collapsing onto the still-forming central star. Incorporation into icy planetesimals could prevent amorphous water ice being destroyed by stellar heating or converted to crystalline form by UV radiation. A large number of frequent collisions are then required to release those icy materials from the planetesimals, converting it to micron-sized dirty ice grains that are detectable using infrared spectroscopy.

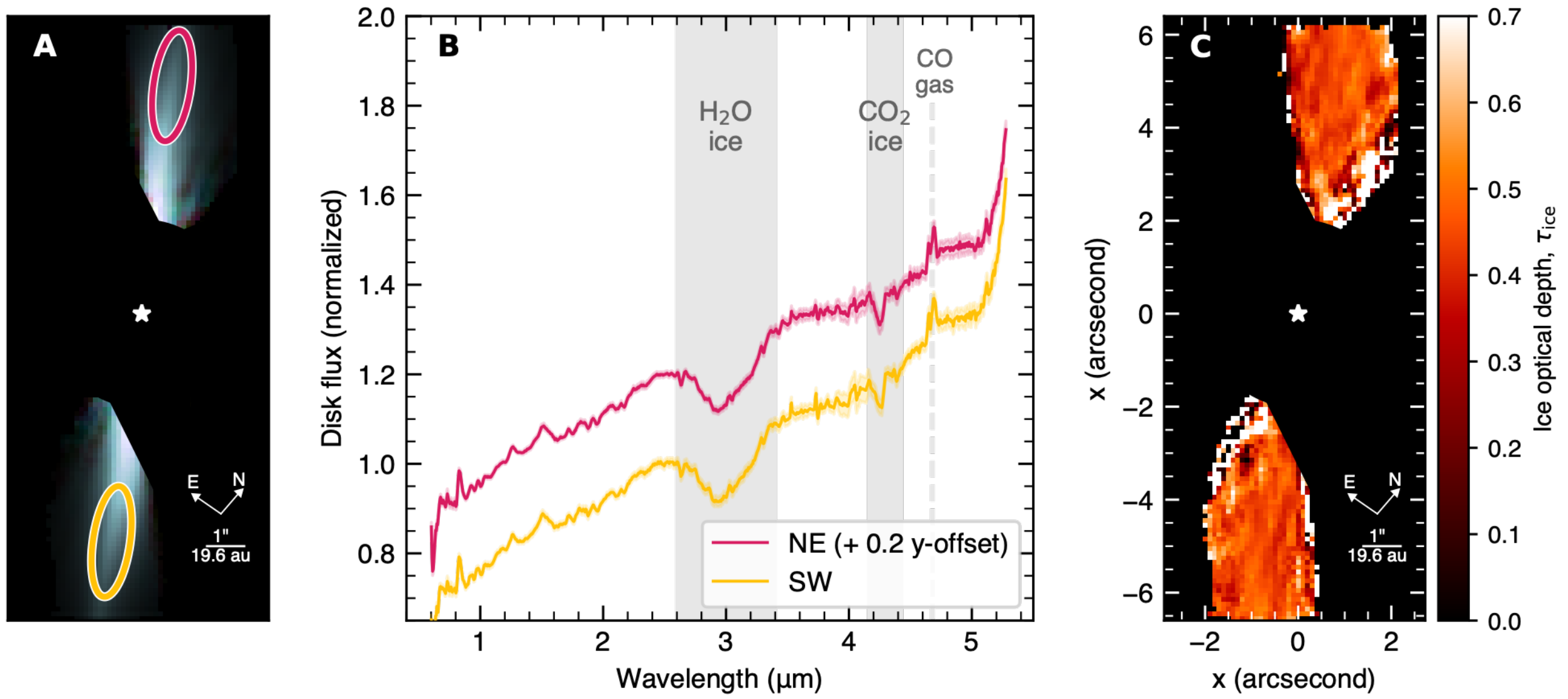

**Fig. 1. Reflectance spectra of the β Pictoris disk.** **(A)** False-color image of the β Pictoris debris disk (located at 05:47:17.087 -51:03:59.441), derived from the NIRSpec IFU data after subtraction of the stellar light. Ovals indicate the NE (magenta) and SW (gold) apertures. The location of the central star is indicated by a white star, and the black area indicates regions that were masked due to residuals from the stellar subtraction process. A continuum component from 2.3 to 2.7 μm, a $H_2O$ ice component from 2.8 to 3.2 μm, and a 0.6 × continuum component from 3.3 to 4.2 μm were mixed to create a cyan-magenta-yellow image (*19*). **(B)** Disk reflectance spectra, normalized at 2.5 μm, extracted from the apertures in panel A, in the same colors. The 1σ uncertainty is shown as light shading. The NE spectrum is offset +0.2 in the y-direction so that the spectra do not overlap. Gray bands are at the expected wavelengths of the $H_2O$ and $CO_2$ ice features, while the vertical dashed gray line indicates the emission due to CO gas. Other narrow features (such as at ~0.9 μm) are due to residual data reduction artifacts (*19*). **(C)** Map of the measured depth of the $H_2O$ ice feature (Equation S5) with the coordinates offset from the location of the central star .

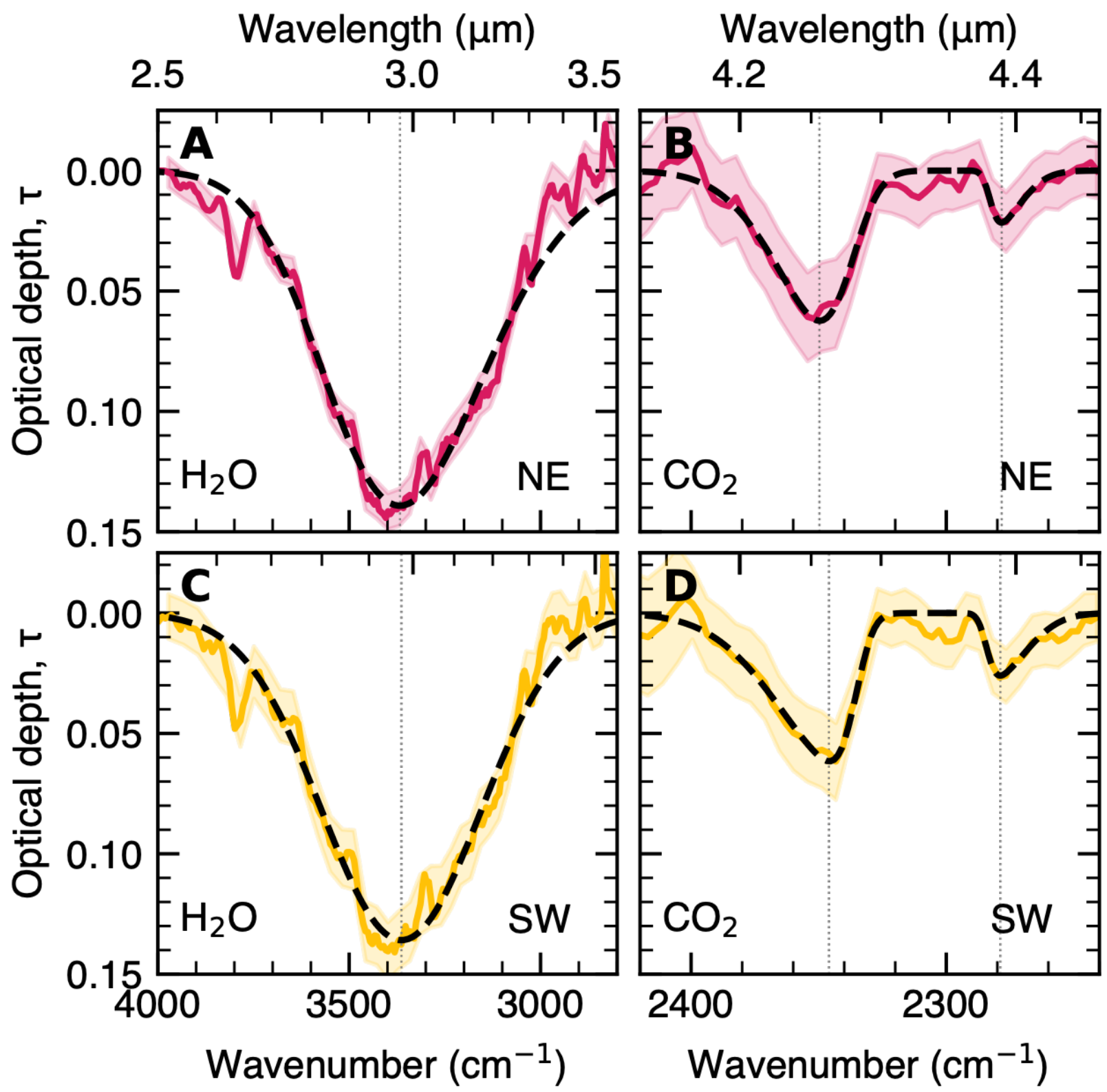

**Fig. 2. Optical depth profiles of the $H_2O$ and $CO_2$ features. (A)** The $H_2O$ feature from the NE aperture (magenta line), after continuum subtraction (Fig. S7) (*19*) and conversion to effective optical depth. The black dashed line is a skewed Gaussian model fitted to the data. **(B)** Same as panel A, but for the $^{12}CO_2$ and $^{13}CO_2$ features. **(C-D)** Same as panels A-B but for the SW aperture (gold line). The colored shading is the 1σ uncertainty on the optical depth.

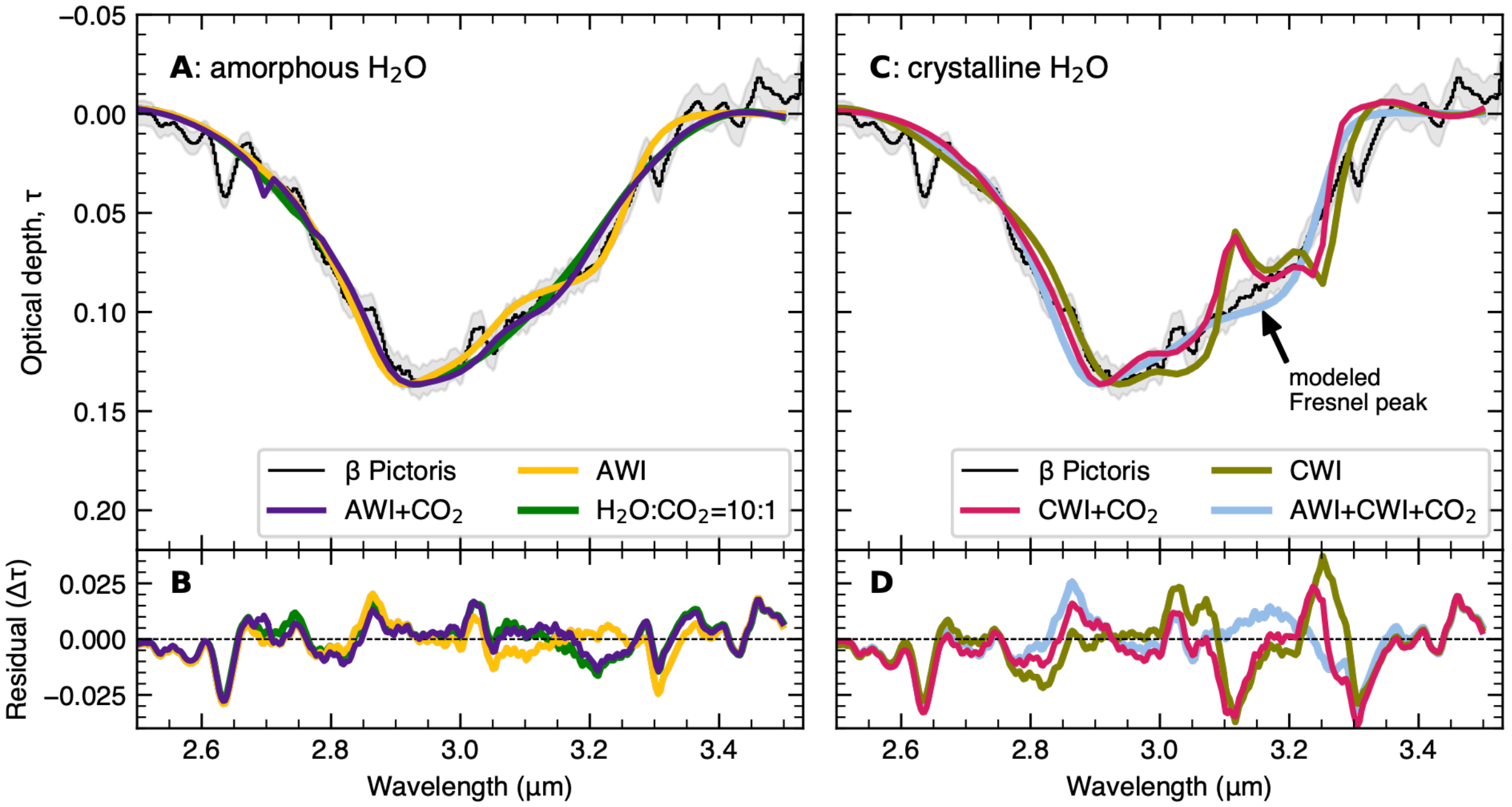

**Fig. 3. $H_2O$ profiles for water ice structures. (A)** The black line is the optical depth of the $H_2O$ feature in the NE aperture with 1σ uncertainty as gray shading (as in Fig. 2A). Colored lines are models of AWI on silicate and carbonaceous dust grains: pure AWI (gold), grains mixed with AWI and $CO_2$ (purple), and grains aggregated with a $H_2O$:$CO_2$=10:1 mixture (green). **(B)** The residuals between the models and the observed spectrum in panel A. The dashed black line is a horizontal reference line at 0 residual. **(C-D)** Same as panels A-B but for models including CWI: pure CWI (olive), grains mixed with CWI and $CO_2$ (magenta), and grains mixed with AWI, CWI, and $CO_2$ (blue).

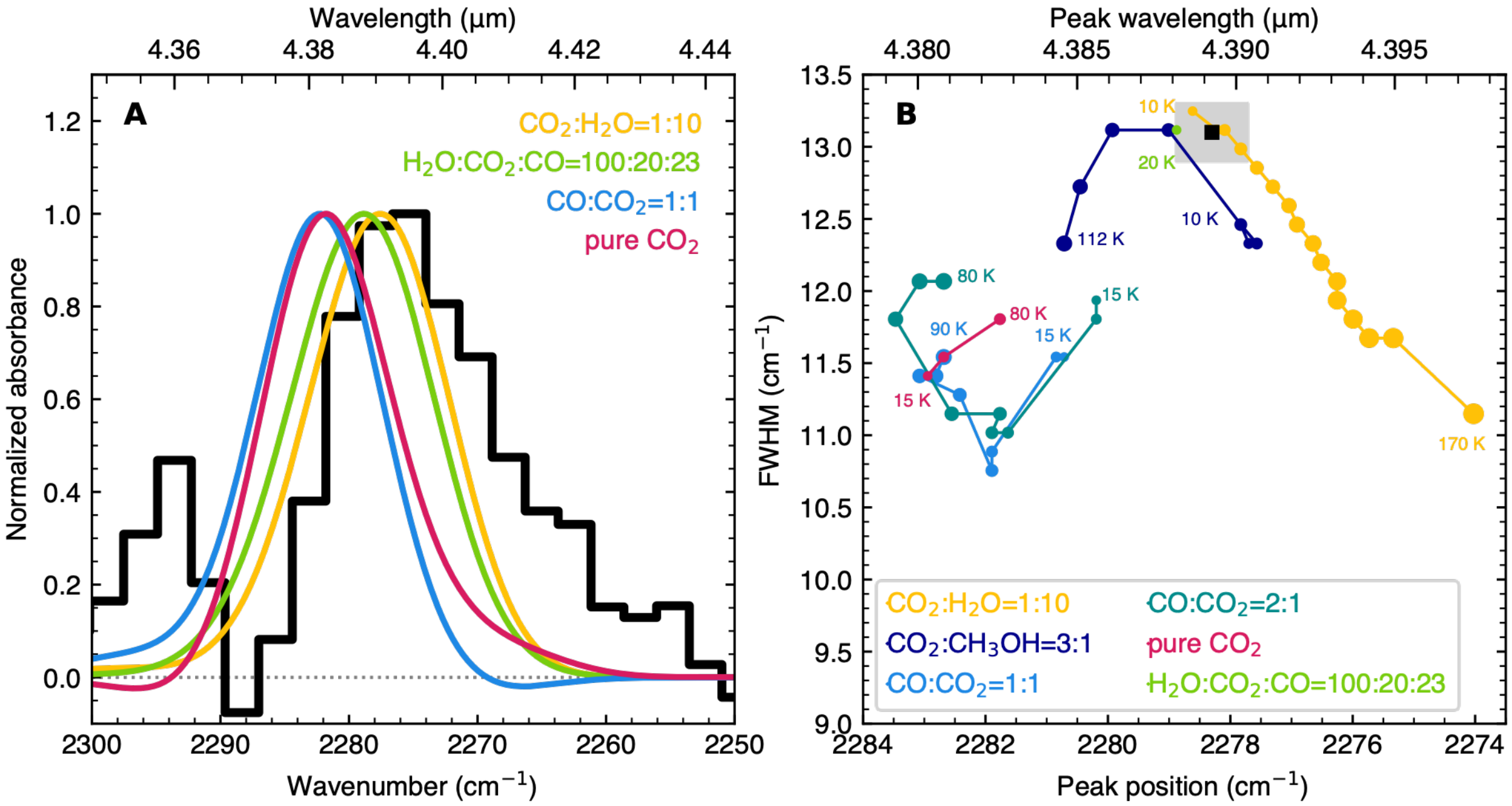

**Fig. 4. Profile shape of the $^{13}CO_2$ band. (A)** The black histogram is the observed $^{13}CO_2$ absorbance in the NE aperture. Colored lines are laboratory spectra of this absorbance in different ice mixtures: $CO:CO_2$=1:1 at 30 K (blue), $H_2O:CO_2:CO$=100:20:23 at 20 K (green), $CO_2:H_2O$=1:10 at 40 K (gold) and pure crystalline $CO_2$ at 80 K (magenta). The spectra are normalized at their maximum absorbance, where absorbance was calculated as $\frac{\text{optical depth}}{\ln(10)} = -\log_{10}(\frac{\text{line depth}}{\text{continuum}})$ . The black dotted line is a horizontal reference line at zero absorbance. **(B)** The FWHM as a function of the peak position for $^{13}CO_2$ feature. The black square is the value measured from the observed spectrum in panel A, with gray shading indicating its 1σ uncertainty. Colored dots are values measured from the laboratory spectra, convolved with the NIRSpec spectral resolution and connected by corresponding colored lines. Dot size increases with the experimental temperature; labels indicate the highest and lowest measured values. Temperature steps between dots differ between mixtures and step, ranging from 3 to 50 K.

**Table 1. Ice optical depth measurements.** Peak wavelength ($\lambda_{\mathrm{peak}}$), peak optical depth ($\tau_{\mathrm{peak}}$), integrated optical depth ($\int \tau d\nu$), half width at half maximum (HWHM), integrated signal-to-noise ratio (S/N$_{\mathrm{int.}}$), effective column densities ($N$), and abundances ($N(CO_2)/N(H_2O)$) are given for $H_2O$, $^{12}CO_2$, and $^{13}CO_2$ for the NE and SW arms. The top values of the HWHM are for the left (lower wavelength) side of the skewed Gaussian model, and the bottom values are for the right side.

| | | $\lambda_{\mathrm{peak}}$ (μm) | $\tau_{\mathrm{peak}}$ | $\int \tau d\nu$ (cm$^{-1}$) | HWHM (cm$^{-1}$) | S/N$_{\mathrm{int.}}$ | $N$ (cm$^{-2}$) | $\frac{N(CO_2)}{N(H_2O)}$ (%) |
|---|---|---|---|---|---|---|---|---|
| NE | $H_2O$ | 2.9697± 0.0101 | 0.1392± 0.0016 | 67.595± 0.088 | 263.40±1.39 231.90±0.84 | 762 | $(3.385\pm0.0041)\times10^{17}$ | 9.21± 0.46 |
| | $^{12}CO_2$ | 4.2558± 0.0022 | 0.0623± 0.0025 | 2.352± 0.088 | 14.92±0.31 22.32±0.37 | 27 | $(3.12\pm0.15)\times10^{16}$ | |
| | $^{13}CO_2$ | 4.3893± 0.0011 | 0.0216± 0.0012 | 0.265± 0.035 | 10.77±0.24 4.959±0.083 | 7 | $(3.41\pm0.17)\times10^{15}$ | |
| SW | $H_2O$ | 2.9736± 0.0101 | 0.1358± 0.0016 | 65.58± 0.14 | 245.30±1.32 242.90±1.09 | 487 | $(3.286\pm0.0041)\times10^{17}$ | 9.42± 0.38 |
| | $^{12}CO_2$ | 4.2625± 0.0015 | 0.0615± 0.0023 | 2.344± 0.076 | 10.84±0.28 26.06±0.37 | 30 | $(3.10\pm0.12)\times10^{16}$ | |
| | $^{13}CO_2$ | 4.3882± 0.0021 | 0.0259± 0.0025 | 0.359± 0.031 | 16.57±0.59 5.904±0.083 | 11 | $(4.61\pm0.28)\times10^{15}$ | |

## References and Notes

1. N. Miret-Roig, P. A. B. Galli, W. Brandner, H. Bouy, D. Barrado, J. Olivares, T. Antoja, M. Romero-Gómez, F. Figueras, J. Lillo-Box, Dynamical traceback age of the β Pictoris moving group. *Astronomy and Astrophysics* **642**, A179 (2020).

2. Gaia Collaboration, A. Vallenari, A. G. A. Brown, T. Prusti, J. H. J. de Bruijne, F. Arenou, C. Babusiaux, M. Biermann, O. L. Creevey, C. Ducourant, D. W. Evans, L. Eyer, R. Guerra, A. Hutton, C. Jordi, S. A. Klioner, U. L. Lammers, L. Lindegren, X. Luri, F. Mignard, C. Panem, D. Pourbaix, S. Randich, P. Sartoretti, C. Soubiran, P. Tanga, N. A. Walton, C. A. L. Bailer-Jones, U. Bastian, R. Drimmel, F. Jansen, D. Katz, M. G. Lattanzi, F. van Leeuwen, J. Bakker, C. Cacciari, J. Castañeda, F. De Angeli, C. Fabricius, M. Fouesneau, Y. Frémat, L. Galluccio, A. Guerrier, U. Heiter, E. Masana, R. Messineo, N. Mowlavi, C. Nicolas, K. Nienartowicz, F. Pailler, P. Panuzzo, F. Riclet, W. Roux, G. M. Seabroke, R. Sordo, F. Thévenin, G. Gracia-Abril, J. Portell, D. Teyssier, M. Altmann, R. Andrae, M. Audard, I. Bellas-Velidis, K. Benson, J. Berthier, R. Blomme, P. W. Burgess, D. Busonero, G. Busso, H. Cánovas, B. Carry, A. Cellino, N. Cheek, G. Clementini, Y. Damerdji, M. Davidson, P. de Teodoro, M. Nuñez Campos, L. Delchambre, A. Dell'Oro, P. Esquej, J. Fernández-Hernández, E. Fraile, D. Garabato, P. García-Lario, E. Gosset, R. Haigron, J. -L. Halbwachs, N. C. Hambly, D. L. Harrison, J. Hernández, D. Hestroffer, S. T. Hodgkin, B. Holl, K. Janßen, G. Jevardat de Fombelle, S. Jordan, A. Krone-Martins, A. C. Lanzafame, W. Löffler, O. Marchal, P. M. Marrese, A. Moitinho, K. Muinonen, P. Osborne, E. Pancino, T. Pauwels, A. Recio-Blanco, C. Reylé, M. Riello, L. Rimoldini, T. Roegiers, J. Rybizki, L. M. Sarro, C. Siopis, M. Smith, A. Sozzetti, E. Utrilla, M. van Leeuwen, U. Abbas, P. Ábrahám, A. Abreu Aramburu, C. Aerts, J. J. Aguado, M. Ajaj, F.

Aldea-Montero, G. Altavilla, M. A. Álvarez, J. Alves, F. Anders, R. I. Anderson, E. Anglada Varela, T. Antoja, D. Baines, S. G. Baker, L. Balaguer-Núñez, E. Balbinot, Z. Balog, C. Barache, D. Barbato, M. Barros, M. A. Barstow, S. Bartolomé, J. -L. Bassilana, N. Bauchet, U. Becciani, M. Bellazzini, A. Berihuete, M. Bernet, S. Bertone, L. Bianchi, A. Binnenfeld, S. Blanco-Cuaresma, A. Blazere, T. Boch, A. Bombrun, D. Bossini, S. Bouquillon, A. Bragaglia, L. Bramante, E. Breedt, A. Bressan, N. Brouillet, E. Brugaletta, B. Bucciarelli, A. Burlacu, A. G. Butkevich, R. Buzzi, E. Caffau, R. Cancelliere, T. Cantat-Gaudin, R. Carballo, T. Carlucci, M. I. Carnerero, J. M. Carrasco, L. Casamiquela, M. Castellani, A. Castro-Ginard, L. Chaoul, P. Charlot, L. Chemin, V. Chiaramida, A. Chiavassa, N. Chornay, G. Comoretto, G. Contursi, W. J. Cooper, T. Cornez, S. Cowell, F. Crifo, M. Cropper, M. Crosta, C. Crowley, C. Dafonte, A. Dapergolas, M. David, P. David, P. de Laverny, F. De Luise, R. De March, Gaia Data Release 3. Summary of the content and survey properties. *Astronomy and Astrophysics* **674**, A1 (2023).

3. M. Cavallius, G. Cataldi, A. Brandeker, G. Olofsson, B. Larsson, R. Liseau, Upper limits on the water vapour content of the β Pictoris debris disk. *Astronomy and Astrophysics* **628**, A127 (2019).

4. I. Rebollido, C. C. Stark, J. Kammerer, M. D. Perrin, K. Lawson, L. Pueyo, C. Chen, D. Hines, J. H. Girard, K. Worthen, C. Ingerbretsen, S. Betti, M. Clampin, D. Golimowski, K. Hoch, N. K. Lewis, C. X. Lu, R. P. van der Marel, E. Rickman, S. Seager, R. Soummer, J. A. Valenti, K. Ward-Duong, C. M. Mountain, JWST-TST High Contrast: Asymmetries, Dust Populations, and Hints of a Collision in the β Pictoris Disk with NIRCam and MIRI. *The Astronomical Journal* **167**, 69 (2024).

5. W. R. F. Dent, M. C. Wyatt, A. Roberge, J. -C. Augereau, S. Casassus, S. Corder, J. S. Greaves, I. de Gregorio-Monsalvo, A. Hales, A. P. Jackson, A. Meredith Hughes, A. -M. Lagrange, B. Matthews, D. Wilner, Molecular Gas Clumps from the Destruction of Icy Bodies in the β Pictoris Debris Disk. *Science* **343**, 1490-1492 (2014).

6. G. Cataldi, A. Brandeker, Y. Wu, C. Chen, W. Dent, B. L. de Vries, I. Kamp, R. Liseau, G. Olofsson, E. Pantin, A. Roberge, ALMA Resolves C I Emission from the β Pictoris Debris Disk. *The Astrophysical Journal* **861**, 72 (2018).

7. Q. Kral, L. Matrà, M. C. Wyatt, G. M. Kennedy, Predictions for the secondary CO, C and O gas content of debris discs from the destruction of volatile-rich planetesimals. *Monthly Notices of the Royal Astronomical Society* **469**, 521-550 (2017).

8. GRAVITY Collaboration, M. Nowak, S. Lacour, P. Mollière, J. Wang, B. Charnay, E. F. van Dishoeck, R. Abuter, A. Amorim, J. P. Berger, H. Beust, M. Bonnefoy, H. Bonnet, W. Brandner, A. Buron, F. Cantalloube, C. Collin, F. Chapron, Y. Clénet, V. Coudé Du Foresto, P. T. de Zeeuw, R. Dembet, J. Dexter, G. Duvert, A. Eckart, F. Eisenhauer, N. M. Förster Schreiber, P. Fédou, R. Garcia Lopez, F. Gao, E. Gendron, R. Genzel, S. Gillessen, F. Haußmann, T. Henning, S. Hippler, Z. Hubert, L. Jocou, P. Kervella, A. -M. Lagrange, V. Lapeyrère, J. -B. Le Bouquin, P. Léna, A. -L. Maire, T. Ott, T. Paumard, C. Paladini, K. Perraut, G. Perrin, L. Pueyo, O. Pfuhl, S. Rabien, C. Rau, G. Rodríguez-Coira, G. Rousset, S. Scheithauer, J. Shangguan, O. Straub, C. Straubmeier, E. Sturm, L. J. Tacconi, F. Vincent, F. Widmann, E. Wieprecht, E. Wiezorrek, J. Woillez, S. Yazici, D. Ziegler, Peering into the formation history of β Pictoris b with VLTI/GRAVITY long-baseline interferometry. *Astronomy and Astrophysics* **633**, A110 (2020).

9. A. Grigorieva, Ph. Thébault, P. Artymowicz, A. Brandeker, Survival of icy grains in debris discs. The role of photosputtering. *Astronomy and Astrophysics* **475**, 755-764 (2007).

10. Y. Wu, K. Worthen, A. Brandeker, C. Chen, Argon in beta Pictoris -- entrapment and release of volatile in disks. *The Astrophysical Journal* **982**, 2 (2025).

11. C. Xie, C. Chen, C. M. Lisse, D. C. Hines, T. Beck, S. K. Betti, N. Pinilla-Alonso, C. Ingebretsen, K. Worthen, A. Gaspar, S. G. Wolff, B. T. Bolin, L. Pueyo, M. D. Perrin, J. A. Stansberry, J. M. Leisenring, Water ice in the debris disk around HD 181327. *Nature* **641**, 608X (2025).

12. C. H. Chen, C. X. Lu, K. Worthen, D. R. Law, B. A. Sargent, A. Moro-Martin, G. C. Sloan, C. M. Lisse, D. M. Watson, J. H. Girard, Y. Chai, D. C. Hines, J. Kammerer, A. Li, M. Perrin, L. Pueyo, I. Rebollido, K. R. Stapelfeldt, C. Stark, M. W. Werner, MIRI MRS Observations of Beta Pictoris. II. The Spectroscopic Case for a Recent Giant Collision. *The Astrophysical Journal* **973**, 139 (2024).

13. S. K. Betti, K. Follette, S. Jorquera, G. Duchêne, J. Mazoyer, M. Bonnefoy, G. Chauvin, L. M. Pérez, A. Boccaletti, C. Pinte, A. J. Weinberger, C. Grady, L. M. Close, D. Defrère, E. C. Downey, P. M. Hinz, F. Ménard, G. Schneider, A. J. Skemer, A. Vaz, Detection of Near-infrared Water Ice at the Surface of the (Pre)Transitional Disk of AB Aur: Informing Icy Grain Abundance, Composition, and Size. *The Astronomical Journal* **163**, 145 (2022).

14. J. A. Sturm, M. K. McClure, T. L. Beck, D. Harsono, J. B. Bergner, E. Dartois, A. C. A. Boogert, J. E. Chiar, M. A. Cordiner, M. N. Drozdovskaya, S. Ioppolo, C. J. Law, H. Linnartz, D. C. Lis, G. J. Melnick, B. A. McGuire, J. A. Noble, K. I. Öberg, M. E. Palumbo, Y. J. Pendleton, G. Perotti, K. M. Pontoppidan, D. Qasim, W. R. M. Rocha, H. Terada, R. G. Urso, E. F. van Dishoeck, A JWST inventory of protoplanetary disk ices. The edge-on protoplanetary disk HH 48 NE, seen with the Ice Age ERS program. *Astronomy and Astrophysics* **679**, A138 (2023).

15. N. P. Ballering, L. Ilsedore Cleeves, R. D. Boyden, M. J. McCaughrean, R. E. Gross, S. G. Pearson, Water Ice in the Edge-on Orion Silhouette Disk 114–426 from JWST NIRCam Images. *The Astrophysical Journal* 979, 110 (2025).

16. A. C. A. Boogert, P. A. Gerakines, D. C. B. Whittet, Observations of the Icy Universe. *The Annual Review of Astronomy and Astrophysics* **53**, 541-581 (2015).

17. P. Jakobsen, P. Ferruit, C. Alves de Oliveira, S. Arribas, G. Bagnasco, R. Barho, T. L. Beck, S. Birkmann, T. Böker, A. J. Bunker, S. Charlot, P. de Jong, G. de Marchi, R. Ehrenwinkler, M. Falcolini, R. Fels, M. Franx, D. Franz, M. Funke, G. Giardino, X. Gnata, W. Holota, K. Honnen, P. L. Jensen, M. Jentsch, T. Johnson, D. Jollet, H. Karl, G. Kling, J. Köhler, M. -G. Kolm, N. Kumari, M. E. Lander, R. Lemke, M. López-Caniego, N. Lützgendorf, R. Maiolino, E. Manjavacas, A. Marston, M. Maschmann, R. Maurer, B. Messerschmidt, S. H. Moseley, P. Mosner, D. B. Mott, J. Muzerolle, N. Pirzkal, J. -F. Pittet, A. Plitzke, W. Posselt, B. Rapp, B. J. Rauscher, T. Rawle, H. -W. Rix, A. Rödel, P. Rumler, E. Sabbi, J. -C. Salvignol, T. Schmid, M. Sirianni, C. Smith, P. Strada, M. te Plate, J. Valenti, T. Wettemann, T. Wiehe, M. Wiesmayer, C. J. Willott, R. Wright, P. Zeidler, C. Zincke, The Near-Infrared Spectrograph (NIRSpec) on the James Webb Space Telescope. I. Overview of the instrument and its capabilities. *Astronomy and Astrophysics* **661**, A80 (2022).

18. T. Böker, S. Arribas, N. Lützgendorf, C. Alves de Oliveira, T. L. Beck, S. Birkmann, A. J. Bunker, S. Charlot, G. de Marchi, P. Ferruit, G. Giardino, P. Jakobsen, N. Kumari, M. López-Caniego, R. Maiolino, E. Manjavacas, A. Marston, S. H. Moseley, J. Muzerolle, P. Ogle, N. Pirzkal, B. Rauscher, T. Rawle, H. -W. Rix, E. Sabbi, B. Sargent, M. Sirianni, M. te Plate, J. Valenti, C. J. Willott, P. Zeidler, The Near-Infrared Spectrograph (NIRSpec) on the James Webb Space Telescope. III. Integral-field spectroscopy. *Astronomy and Astrophysics* **661**, A82 (2022).

19. Materials and methods are available as supplementary materials.

20. A. Roberge, P. D. Feldman, A. M. Lagrange, A. Vidal-Madjar, R. Ferlet, A. Jolly, J. L. Lemaire, F. Rostas, High-Resolution HUBBLE SPACE TELESCOPE STIS Spectra of C I and CO in the β Pictoris Circumstellar Disk. *The Astrophysical Journal* **538**, 904 (2000).

21. M. R. Troutman, K. H. Hinkle, J. R. Najita, T. W. Rettig, S. D. Brittain, Ro-vibrational CO Detected in the β Pictoris Circumstellar Disk. *The Astrophysical Journal* **738**, 12 (2011).

22. S. A. Sandford, L. J. Allamandola, The Physical and Infrared Spectral Properties of CO 2 in Astrophysical Ice Analogs. *The Astrophysical Journal* 355, 357 (1990).

23. K. Isokoski, C. A. Poteet, and H. Linnartz, Highly resolved infrared spectra of pure $CO_2$ ice (15-75 K). *Astronomy and Astrophysics* **555**, A85 (2013).

24. A. C. A. Boogert, P. Ehrenfreund, P. A. Gerakines, A. G. G. M. Tielens, D. C. B. Whittet, W. A. Schutte, E. F. van Dishoeck, Th. de Graauw, L. Decin, T. Prusti, ISO-SWS observations of interstellar solid $^{13}CO_2$: heated ice and the Galactic $^{12}C/^{13}C$ abundance ratio. *Astronomy and Astrophysics* **353**, 349-362 (2000).

25. D. Bockelée-Morvan, N. Biver, C. A. Schambeau, J. Crovisier, C. Opitom, M. de Val-Borro, E. Lellouch, P. Hartogh, B. Vandenbussche, E. Jehin, M. Kidger, M. Küppers, D. C. Lis, R. Moreno, S. Szutowicz, V. Zakharov, Water, hydrogen cyanide, carbon monoxide, and dust production from distant comet 29P/Schwassmann-Wachmann 1. *Astronomy and Astrophysics* **664**, A95 (2022).

26. M. J. Mumma, S. B. Charnley, The Chemical Composition of Comets—Emerging Taxonomies and Natal Heritage. *Annual Review of Astronomy and Astrophysics* 49, 471-524 (2011).

27. M. K. McClure, M. van't Hoff, L. Francis, E. Bergin, W. R. M. Rocha, J. A. Sturm, D. Harsono, E. F. van Dishoeck, J. H. Black, J. A. Noble, D. Qasim, E. Dartois, Refractory solid condensation detected in an embedded protoplanetary disk. *Nature* 643, 649-653 (2025).

28. N. G. C. Brunken, W. R. M. Rocha, E. F. van Dishoeck, R. Gutermuth, H. Tyagi, K. Slavicinska, P. Nazari, S. T. Megeath, N. J. Evans, M. Narang, P. Manoj, A. E. Rubinstein, D. M. Watson, L. W. Looney, H. Linnartz, A. Caratti o Garatti, H. Beuther, H. Linz, P. Klaassen, C. A. Poteet, S. Federman, G. Anglada, P. Atnagulov, T. L. Bourke, W. J. Fischer, E. Furlan, J. Green, N. Habel, L. Hartmann, N. Karnath, M. Osorio, J. Muzerolle Page, R. Pokhrel, R. Rahatgaonkar, P. Sheehan, T. Stanke, A. M. Stutz, J. J. Tobin, L. Tychoniec, S. Wolk, Y-L. Yang, JWST observations of $^{13}CO_2$ ice. Tracing the chemical environment and thermal history of ices in protostellar envelopes. *Astronomy and Astrophysics* **685**, A27 (2024).

29. R. M. Mastrapa, S. A. Sandford, T. L. Roush, D. P. Cruikshank, and C. M. Dalle Ore, Optical Constants of Amorphous and Crystalline $H_2O$-ice: 2.5-22 μm (4000-455 $cm^{-1}$) Optical Constants of $H_2O$-ice. *The Astrophysical Journal* 701, 1347-1356 (2009).

30. W. Hagen, R. M. Humphreys, R. E. Stencel, High-dispersion spectroscopy of the most luminous F-and G- type supergiants in the Large Magellanic Cloud and the milky way. *Publications of the Astronomical Society of the Pacific* 93, 567-576 (1981).

31. C. Reinert, H. Mutschke, A. V. Krivov, T. Löhne, P. Mohr, Absorption of crystalline water ice in the far infrared at different temperatures. *Astronomy and Astrophysics* **573**, A29 (2015).

32. E. F. van Dishoeck, E. Herbst, D. A. Neufeld, Interstellar Water Chemistry: From Laboratory to Observations. *Chemical Reviews* **113**, 9043-9085 (2013).

33. C. M. Lisse, J. K. Steckloff, D. Prialnik, M. Womack, O. Harrington Pinto, G. Sarid, Y. R. Fernandez, C. A. Schambeau, T. Kareta, N. H. Samarasinha, W. Harris, K. Volk, L. M. Woodney, D. P. Cruikshank, and S. A. Sandford, 29P/Schwassmann-Wachmann 1: A Rosetta Stone for Amorphous Water Ice and CO ↔ $CO_2$ Conversion in Centaurs and Comets?. *The Planetary Science Journal* **3**, 251 (2022).

34. R. G. Smith, G. Robinson, A. R. Hyland, G. L. Carpenter, Molecular ices as temperature indicators for interstellar dust: the 44- and 62-μm lattice features of $H_2O$ ice. *Monthly Notices of the Royal Astronomical Society* **271**, 481-489 (1994).

35. M. Famá, M. J. Loeffler, U. Raut, R. A. Baragiola, Radiation-induced amorphization of crystalline ice. *Icarus* **207**, 314-319 (2010).

36. K. M. Pontoppidan, A. C. A. Boogert, H. J. Fraser, E. F. van Dishoeck, G. A. Blake, F. Lahuis, K. I. Öberg, N. J. Evans, C. Salyk, The c2d Spitzer Spectroscopic Survey of Ices around Low-Mass Young Stellar Objects. II. $CO_2$. *The Astrophysical Journal* **678**, 1005-1031 (2008).

37. R. M. Escribano, G. M. Muñoz Caro, G. A. Cruz-Diaz, Y. Rodríguez-Lazcano, B. Maté, Crystallization of CO2 ice and the absence of amorphous CO2 ice in space, *Proceedings of the National Academy of Sciences of the United States of America* **110**, 12899-12904 (2013).

38. A. Potapov, H. Linz, J. Bouwman, W. R. M. Rocha, J. S. Martin, S. Wolf, Th. Henning, H. Terada, Simple molecules and complex chemistry in a protoplanetary disk: A JWST investigation of the highly inclined disk d216-0939. *Astronomy and Astrophysics* 697, A53 (2025).

39. K. Worthen, C. H. Chen, D. R. Law, C. X. Lu, K. Hoch, Y. Chai, G. C. Sloan, B. A. Sargent, J. Kammerer, D. C. Hines, I. Rebollido, W. Balmer, M. D. Perrin, D. M. Watson, L. Pueyo, J. H. Girard, C. M. Lisse, C. C. Stark, MIRI MRS Observations of β Pictoris. I. The Inner Dust, the Planet, and the Gas. *The Astrophysical Journal* **964**, 168 (2024).

40. J. He, SM Emtiaz, A. Boogert, G. Vidali, The $^{12}CO_2$ and $^{13}CO_2$ Absorption Bands as Tracers of the Thermal History of Interstellar Icy Grain Mantles. *The Astrophysical Journal* **869**, 41 (2018).

41. L. E. Strubbe, E. I. Chiang, Dust Dynamics, Surface Brightness Profiles, and Thermal Spectra of Debris Disks: The Case of AU Microscopii. *The Astrophysical Journal* **648**, 652-665 (2006).

42. J. K. Steckloff, G. Sarid, B. C. Johnson, The Effects of Early Collisional Evolution on Amorphous Water Ice Bodies. *The Planetary Science Journal* **4**, 4 (2023).

43. Barbara A. Mikulski Archive for Space Telescopes, 1564 B Pic observations, MAST archive (2026); https://dx.doi.org/10.17909/sdcd-a118

44. Barbara A. Mikulski Archive for Space Telescopes, Program 1128 data, MAST archive (2026); https://dx.doi.org/10.17909/xry0-6496

45. S. Betti, C. H. Chen, C. Xie, K. Worthen, S. Brittain, J. Najita, I. Rebollido, T. Beck, B. Bolin, D. Hines, C. Ingebretsen, C. Lisse, M. D. Perrin, L. Pueyo, I. Wong, C. Stark, Detection of Ices in the Debris Disk around β Pictoris, version 1, Dryad (submitted); https://doi.org/10.5061/dryad.n8pk0p36d

46. H. Bushouse, J. Eisenhamer, N. Dencheva, J. Davies, P, Greenfield, J. Morrison, P. Hodge, B. Simon, D. Grumm, M. Droettboom, E. Slavich, M. Sosey, T. Pauly, T. Miller, R. Jedrzejewski, W. Hack, D. Davis, S. Crawford, D. Law, K. Gordon, M. Regan, M. Cara, K. MacDonald, L. Bradley, C. Shanahan, W. Jamieson, M. Teodoro, T. Williams, M. Pena-Guerrero, B. Graham, E. Molter, T. Brandt, C. Hayes, R. Cooper, M. Clarke, J. Filippazzo, JWST Calibration Pipeline. Zenodo (2025). https://doi.org/10.5281/zenodo.21364698

47. P. Greenfield,T. Miller, The Calibration Reference Data System. *Astronomy and Computing* **16**, 41-53 (2016).

48. B. Ren, J. J. Wang, L. Pueyo, centerRadon: Center determination code in stellar images, GitHub (2019); https://github.com/seawander/centerRadon

49. L. Pueyo, R. Soummer, J. Hoffmann, R. Oppenheimer, J. R. Graham, N. Zimmerman, C. Zhai, J. K. Wallace, F. Vescelus, A. Veicht, G. Vasisht, T. Truong, A. Sivaramakrishnan, M. Shao, L. C. Roberts, J. E. Roberts, E. Rice, I. R. Parry, R. Nilsson, T. Lockhart, E. R. Ligon, D. King, S. Hinkley, L. Hillenbrand, D. Hale, R. Dekany, J. R. Crepp, E. Cady, R. Burruss, D. Brenner, C. Beichman, C. Baranec, Reconnaissance of the HR 8799 Exosolar System. II. Astrometry and Orbital Motion. *The Astrophysical Journal* **803**, 31 (2015).

50. B. Ren, M. Perrin, seawander/DebrisDiskFM: First Release, Zenodo (2018); https://doi.org/10.5281/zenodo.2398963

51. L. G. Henyey, J. L. Greenstein, Diffuse radiation in the Galaxy. *The Astrophysical Journal* **93**, 70-83 (1941).

52. M. Janson, A. Brandeker, G. Olofsson, R. Liseau, The far reaches of the β Pictoris debris disk. *Astronomy and Astrophysics* **646**, A132 (2021).

53. M. Ahmic, B. Croll, P. Artymowicz, Dust Distribution in the β Pictoris Circumstellar Disks. *The Astrophysical Journal* **705**, 529-542 (2009).

54. J. Milli, A. -M. Lagrange, D. Mawet, O. Absil, J. -C. Augereau, D. Mouillet, A. Boccaletti, J. H. Girard, G. Chauvin, Very deep images of the innermost regions of the β Pictoris debris disc at L'. *Astronomy and Astrophysics* **566**, A91 (2014).

55. R. Storm, K. Price, Differential Evolution - A Simple and Efficient Heuristic for global Optimization over Continuous Spaces. *Journal of Global Optimization* **11**, 341-359 (1997).

56. K. Lawson, T. Currie, J. P. Wisniewski, M. Tamura, G. Schneider, J-C. Augereau, T. D. Brandt, O. Guyon, N. J. Kasdin, T. D. Groff, J. Lozi, J. Chilcote, K. Hodapp, N. Jovanovic, F. Martinache, N. Skaf, E. Akiyama, T. Henning, G. R. Knapp, J. Kwon, S. Mayama, M. W. McElwain, M. L. Sitko, R. Asensio-Torres, T. Uyama, K. Wagner, SCExAO/CHARIS Near-infrared Integral Field Spectroscopy of the HD 15115 Debris Disk. *The Astronomical Journal* **160**, 163 (2020).

57. C. X. Lu, C. H. Chen, B. A. Sargent, D. M. Watson, C. M. Lisse, J. D. Green, M. L. Sitko, T. Mittal, V. Lebouteiller, G. C. Sloan, I. Rebollido, D. C. Hines, J. H. Girard, M. W. Werner, K. R. Stapelfeldt, W. Wu, K. Worthen, Trends in Silicates in the β Pictoris Disk. *The Astrophysical Journal* **933**, 54 (2022).

58. E. Dartois, J. A. Noble, M. K. McClure, J. A. Sturm, T. L. Beck, N. Arulanantham, M. N. Drozdovskaya, C. C. Espaillat, D. Harsono, M. -E. Palumbo, Y. J. Pendleton, K. M. Pontoppidan, The edge-on disc Tau 042021: Icy grains at high altitudes and a wind containing astronomical polycyclic aromatic hydrocarbons. *Astronomy and Astrophysics* 698, A8 (2025).

59. P. A. Gerakines, W. A. Schutte, J. M. Greenberg, E. F. van Dishoeck, The infrared band strengths of $H_2O$, CO and $CO_2$ in laboratory simulations of astrophysical ice mixtures. *Astronomy and Astrophysics* 296, 810 (1995).

60. M. Newville, R. Otten, A. Nelson, T. Stensitzki, A. Ingargiola, D. Allan, A. Fox, F. Carter, Michał, R. Osborn, D. Pustakhod, S. Weigand, lneuhaus, A. Aristov, Glenn, Mark, mgunyho, C. Deil, A. L. R. Hansen, G. Pasquevich, L. Foks, N. Zobrist, O. Frost, Stuermer, J.-C. Jaskula, S. Caldwell, P. Eendebak, M. Pompili, J. H. Nielsen, A. Persaud, lmfit/lmfit-py, version 1.3.2, Zenodo (2024); https://doi.org/10.5281/zenodo.12785036.

61. S. Prahl, Miepython: Pure Python Calculation of Mie Scattering. V. 3.3.0. Zenodo, (2026); https://doi.org/10.5281/zenodo.21716559

62. J. R. Graham, P. G. Kalas, B. C. Matthews, The Signature of Primordial Grain Growth in the Polarized Light of the AU Microscopii Debris Disk. *The Astrophysical Journal* **654**, 595-605 (2007).

63. M. D. Perrin, G. Duchene, M. Millar-Blanchaer, M. P. Fitzgerald, J. R. Graham, S. J. Wiktorowicz, P. G. Kalas, B. Macintosh, B. Bauman, A. Cardwell, J. Chilcote, R. J. De Rosa, D. Dillon, R. Doyon, J. Dunn, D. Erikson, D. Gavel, S. Goodsell, M. Hartung, P. Hibon, P. Ingraham, D. Kerley, Q. Konapacky, J. E. Larkin, J. Maire, F. Marchis, C. Marois, T. Mittal, K. M. Morzinski, B. R. Oppenheimer, D. W. Palmer, J. Patience, L. Poyneer, L. Pueyo, F. T. Rantakyrö, N. Sadakuni, L. Saddlemyer, D. Savransky, R. Soummer, A. Sivaramakrishnan, I. Song, S. Thomas, J. K. Wallace, J. J. Wang, S. G. Wolff, Polarimetry with the Gemini Planet Imager: Methods, Performance at First Light, and the Circumstellar Ring around HR 4796A. *The Astrophysical Journal* **799**, 182 (2015).

64. P. Arriaga, M. P. Fitzgerald, G. Duchêne, P. Kalas, M. A. Millar-Blanchaer, M. D. Perrin, C. H. Chen, J. Mazoyer, M. Ammons, V. P. Bailey, T. S. Barman, J. Bulger, J. K. Chilcote, T. Cotten, R. J. De Rosa, R. Doyon, T. M. Esposito, K. B. Follette, B. L. Gerard, S. Goodsell, J. R. Graham, A. Z. Greenbaum, P. Hibon, J. Hom, L.-W. Hung, P. Ingraham, Q. M. Konopacky, B. A. Macintosh, J. Maire, F. Marchis, M. S. Marley, C. Marois, S.

Metchev, E. L. Nielsen, R. Oppenheimer, D. W. Palmer, J. Patience, L. A. Poyneer, L. Pueyo, A. Rajan, J. Rameau, F. T. Rantakyrö, J.-B. Ruffio, D. Savransky, A. C. Schneider, A. Sivaramakrishnan, I. Song, R. Soummer, S. Thomas, J. J. Wang, K. Ward-Duong, S. G. Wolff, Multiband Polarimetric Imaging of HR 4796A with the Gemini Planet Imager. *The Astronomical Journal* **160**, 79 (2020).

65. D. A. G. Bruggeman, Berechnung verschiedener physikalischer Konstanten von heterogenen Substanzen. I. Dielektrizitätskonstanten und Leitfähigkeiten der Mischkörper aus isotropen Substanzen. *Annalen der Physik* **416**, 636–664 (1935)
66. J. Dorschner, B. Begemann, T. Henning, C. Jaeger, H. Mutschke, Steps toward interstellar silicate mineralogy. II. Study of Mg-Fe-silicate glasses of variable composition. *Astronomy and Astrophysics* **300**, 503 (1995).
67. V. G. Zubko, V. Mennella, L. Colangeli, E. Bussoletti, Optical constants of cosmic carbon analogue grains - I. Simulation of clustering by a modified continuous distribution of ellipsoids. *Monthly Notices of the Royal Astronomical Society* 282, 1321-1329 (1996).
68. N. P. Ballering, K. Y. L. Su, G. H. Rieke, A. Gáspár, A Comprehensive Dust Model Applied to the Resolved Beta Pictoris Debris Disk from Optical to Radio Wavelengths. *The Astrophysical Journal* 823, 108 (2016).
69. S. G. Warren, Optical constants of carbon dioxide ice. *Applied Optics* 25, 2650-2674 (1986).
70. C. Dominik, M. Min, R. Tazaki, OpTool: Command-line driven tool for creating complex dust opacities. Astrophysics Source Code Library ascl:2104.010 (2021).
71. P. Ehrenfreund, O. Kerkhof, W. A. Schutte, A. C. A. Boogert, P. A. Gerakines, E. Dartois, L. D'Hendecourt, A. G. G. M. Tielens, E. F. van Dishoeck, D. C. B. Whittet, Laboratory studies of thermally processed $H_2O$-$CH_3OH$-$CO_2$ ice mixtures and their astrophysical implications. *Astronomy and Astrophysics* **350**, 240-253 (1999).
72. P. A. Gerakines R. L. Hudson, A Modified Algorithm and Open-source Computational Package for the Determination of Infrared Optical Constants Relevant to Astrophysics. *The Astrophysical Journal* 901, 52 (2020).
73. W. R. M. Rocha, M. G. Rachid, B. Olsthoorn, E. F. van Dishoeck, M. K. McClure, H. Linnartz, LIDA: The Leiden Ice Database for Astrochemistry. *Astronomy and Astrophysics* **668**, A63 (2022).
74. F. A. van Broekhuizen, “A Laboratory Route to Interstellar Ice”, thesis, Leiden University (2005) https://hdl.handle.net/1887/2710.
75. N. Skaf, A. Boccaletti, E. Pantin, P. Thebault, Q. Kral, C. Danielski, R. Galicher, J. Milli, A.-M. Lagrange, C. Baruteau, M. Kenworthy, O. Absil, M. Langlois, J. Olofsson, G. Chauvin, N. Huelamo, P. Delorme, B. Charnay, O. Guyon, M. Bonnefoy, F. Cantalloube, H. J. Hoeijmakers, U. Käufl, M. Kasper, A.-L. Maire, M. Mâlin, R. Siebenmorgen, I. Snellen, G. Zins, The β Pictoris system: Setting constraints on the planet and the disk structures at mid-IR wavelengths with NEAR. *Astronomy and Astrophysics* **675**, A35 (2023) .

**Acknowledgments:** This work is based on observations made with the NASA/ESA/CSA James Webb Space Telescope (JWST). The data were obtained from the Mikulski Archive for Space Telescopes (MAST) at the Space Telescope Science Institute, which is operated by the Association of Universities for Research in Astronomy under NASA contract NAS 5-03127 for JWST.

**Funding:** S.K.B was supported by the STScI Postdoctoral Fellowship. K.W was supported by NASA FINESST grant number 80NSSC22K1752, and by the STScI Director's Research fund. C. X was supported by NASA JWST contract NAS 5-03127.

**Author contributions:** S.K.B carried out the analysis, visualizations, and wrote the paper. C.H.C led the JWST proposal, contributed to the methodology and analysis, and helped write the manuscript. C.X. contributed to the methodology and analysis, wrote the general data analysis and modeling scripts that S.K.B restructured to analyze the object. K.W and C.I used the JWST pipeline to perform the initial data reduction and provided substantial input in the formal analysis and discussion of the visualization. I.R. assisted in the data calibration, helping to make sure the JWST pipeline output was robust. T.B. assisted in the data reduction, provided the PSF data cube, and helped with the PSF convolution of the models. S.B. and J.N assisted in the analysis and interpretation of the data, particularly the column densities, and gave substantial input in the discussion of the visualizations. B.B., C.L, I.W. assisted with the analysis, visualization conception, and provided comparisons to Solar System bodies. T.B., C.L., M.D.P. , L.P. C.S contributed to the planning and writing of the JWST proposal. C.H.C, C.X., K.W., S.B., J.N., I.R., B.B, D.H., C.L, M.D.P., L.P., I.W. made substantial contributions to the interpretation of the results. All authors participated in the discussion and interpretation of the results and made substantial contributions commenting and revising the paper.

**Competing interests:** The authors declare that they have no competing interests.

**Data and materials availability:** The JWST observations are available from MAST https://mast.stsci.edu under program 1563 for β Pictoris and the PSF reference star (*43*), and 1128 for the PSF reference (*44*) used in the disk models. Our data reduction scripts are provided in data S1-S4. Our reduced data products are archived at Dryad (*45*).

**Supplementary Materials**

Materials and Methods

Figs. S1 to S14

Table S1

References (*46–75*)

Data S1 to S4

Science AAAS

# Supplementary Materials for

## Carbon dioxide and amorphous water ices in the debris disk around β Pictoris

S. K. Betti[1*], C. H. Chen[1], C. Xie[2], K. Worthen[2], S. Brittain[3], J. Najita[4], I. Rebollido[5], T. Beck[1], B. T. Bolin[6], D. Hines[1], C. Ingebretsen[2], C. Lisse[7], M. D. Perrin[1], L. Pueyo[1], I. Wong[1], and C. Stark[8]

*Corresponding author: sbetti@stsci.edu

**The PDF file includes:**

Materials and Methods
Figs. S1 to S14
Table S1
References (*46*–*75*)

**Other Supplementary Materials for this manuscript:**

Data S1-S4

## Materials and Methods

JWST observations

β Pictoris was observed on 2023 December 7 with the NIRSpec IFU (*17*, *18*) on JWST as part of program 1563 (*43*). The target was observed using the PRISM/CLEAR configuration providing an average resolving power $R$~100 over the 0.6 to 5.2 μm wavelength range. The NIRSpec (NRS) Improved Reference Sampling and Subtraction (IRS2) RAPID (NRSIRS2RAPID) read-out pattern was used with a nine-point dither pattern and five groups per integration. To cover the majority of the 20.0″ edge-on disk and mitigate saturation from the luminous host star, we positioned four mosaic tiles, two on each side of the central star, such that the central star was not observed. This produced a final mosaic IFU field-of-view (FOV) of 5.6″ × 12.8″. The total on-source exposure time for β Pictoris was 1351.19 s (52.52 min). The star α Pictoris (also catalogued as HD 50241) was observed as a PSF reference, in a non-interruptible sequence using the same observational setup and exposure time.

Data reduction and PSF subtraction

We reduced the β Pictoris and the PSF calibrator using identical methods. After downloading the raw uncalibrated images for each target, we ran them through stages 1 to 3 of the JWST Science Calibration pipeline (v1.14.0) (*46*). We use the Calibration Reference Data System context version jwst_1193.pmap to calibrate the data for each stage of the reduction (*47*). In stage 1, we use the Detector1Pipeline module to correct for bad pixels, saturation, linearity, persistence, dark current, charge migration, pink (1/f) noise, and jump noise. This converted the raw exposures to uncalibrated ramp images. In stage 2, we calibrated the individual ramps using the Spec2Pipeline module, which produced flux calibrated exposures; we excluded background subtraction and 1D spectrum extraction from the default parameters. In stage 3, we used the Spec3Pipeline.call module to produce an IFU data cube with a spatial pixel (spaxel) scale of 0.1″, by combining the dither positions and mosaic tiles. The cube uses detector coordinates to avoid interpolation noise that would be introduced if we converted to sky coordinates (*11*).

In the flux calibrated data cube, we detect the faint disk around β Pictoris. To extract the disk spectra, we perform reference differential imaging (RDI) PSF subtraction to remove the stellar flux from the PSF wings. The script for the PSF subtraction and throughput correction is available in Data S1. Because target acquisition exposures were not performed for these observations, we cannot guarantee that the expected PSF position is identical between the β Pictoris and α Pictoris observations. The data cubes were re-centered using the diffraction spikes in the mosaicked images (*11*). We used the radonCenter software (*48*) which implements the Radon transform centering method (*49*) to determine the stellar position by integrating the image intensity along the direction of each diffraction spike. The star's position was determined by the maximum of the sum of the line integrals. This method constrains the star position with a precision of 0.1 pixels. We then aligned the data cubes to the central star position.

After alignment, we performed PSF subtraction. We masked the diffraction spikes and disk at every wavelength slice, for both β Pictoris and α Pictoris, then scaled each slice of the PSF data cube to minimize the residuals between the β Pictoris and α Pictoris data cubes in the non-masked regions, following previous methods (*11*). The diffraction spikes were masked because their flux does not correspond to the stellar photosphere model, and the disk was masked to prevent over-subtraction when scaling the reference image to the science image in each spectral slice. We minimize the cost function $C$:

$$\arg\min_{\lambda,i} \mathrm{C} = \log\left(\Sigma_{i=1}^{N_{\mathrm{pix}}}\left((I_i - f_{\mathrm{RDI}}R_i)M_i\right)^2\right) \tag{S1}$$

where $\lambda$ is the wavelength slice, $i$ is the pixel index, $N_{\mathrm{pix}}$ is the total number of pixels, $I_i$ is the science image, $f_{\mathrm{RDI}}$ is the scaling factor between the reference and science data cubes used for RDI, $R_i$ is the PSF reference image, and $M_i$ is the mask. We found that the average scaling factor across the whole wavelength range is $\langle f_{\mathrm{RDI}}\rangle = 0.535 \pm 0.020$, with small scale fluctuations. To remove artifact-imposed fluctuations when performing RDI, we apply the average constant scaling factor for every wavelength slice, as $I_i - \langle f_{\mathrm{RDI}}\rangle R_{\mathrm{i}}$. This reduces any instrumental artifacts that may cause an incorrect RDI scaling factor (hot or bad pixels). We show the reduced β Pictoris and α Pictoris cubes and the PSF subtracted cube of β Pictoris in Fig. S1.

Throughput correction and spectral extraction

We then corrected the data for flux loss introduced by any PSF mismatch or over subtraction. Throughput correction is typically performed by injecting a model of the disk at a different position angle (PA) and performing RDI to determine the amount of flux lost in the model during the PSF subtraction. However, the FOV of the NIRSpec observations is too narrow to inject a β Pictoris model. Therefore, we adopt a different technique, known as negative disk injection (*11*).

We first use the DebrisDiskFM package (*50*) to calculate a geometric disk model of β Pictoris, to fit the disk surface brightness. We assume a density distribution of dust particles in cylindrical coordinates:

$$\rho(r,z) = \rho_0[(r/r_c)^{-2\alpha_{\mathrm{in}}} + (r/r_c)^{-2\alpha_{\mathrm{out}}}]^{-1/2}e^{-(z/hr^{\beta})^2} \tag{S2}$$

where $\rho_0$ is the scaling factor, $r_c$ is the critical radius of the disk, $\alpha_{\mathrm{in}}$ and $\alpha_{\mathrm{out}}$ are the rising and decreasing components of an asymptotic power-law index, $h$ is the scale height, and $\beta$ is the flaring index. We approximate the scattering phase function as the Henyey-Greenstein (HG) function (***Error! Reference source not found.***), and tested and then adopted $\alpha_{\mathrm{in}} = 0.5$, $\alpha_{\mathrm{out}} = -1.5$, and $\beta = 1$ (*52*, *53*, *54*). We vary the HG asymmetry parameter $g$ and use a free amplitude scaling parameter $A_{\mathrm{f}}$ to match the model to the disk flux. To convolve the disk models to the NIRSpec PSF, we use an empirical NIRSpec/PRISM PSF observation of 2MASS J16194609+5534178 (also catalogued as UCAC3 292-117285). We used differential evolution optimization (DE) (*55*) to fit this model in each wavelength slice. DE explores multi-dimensional parameter space to find a global minimum (*13*, *56*) by initializing a population of models across the whole parameter space, which are then iteratively evolved towards the best-fitting global solution. We initialize 20 models (10× the number of free parameters) which are scaled (using $A$), convolved, and then subtracted from the science image (negative injection). We ran each model through RDI to find the best fitting model ($f_{\mathrm{RDI,mod}}$) that minimizes the residuals in the regions of the negatively injected disk. We find an average $\langle f_{\mathrm{RDI,mod}}\rangle = 0.38 \pm 0.03$, with fluctuations across the wavelength range (Fig. S2).

To compute the throughput, we use the average $\langle f_{\mathrm{RDI,mod}}\rangle$ and re-run the negative disk injection DE optimization, assuming this constant scaling factor when performing the PSF subtraction. Then, we computed the throughput:

$$T_\lambda = \frac{F_{\mathrm{disk}} - F_{\mathrm{res}}}{F_{\mathrm{model}}} \tag{S3}$$

where $T_\lambda$ is the throughput at each spectral slice, $F_{\mathrm{disk}}$ is the science data cube after PSF subtraction assuming a constant $f_{\mathrm{RDI}}$, $F_{\mathrm{res}}$ is the disk free data cube after performing the negative disk injection assuming a constant $\langle f_{\mathrm{RDI,mod}} \rangle$, and $F_{\mathrm{model}}$ is the best fitting model assuming a constant $\langle f_{\mathrm{RDI,mod}} \rangle$. This throughput was calculated when extracting the spectra using aperture photometry. We determined the throughput at each wavelength slice; to reduce contamination from PSF subtraction, we then averaged the throughput to remove any wavelength dependent fluctuations. A throughput value of 1 indicated no flux loss, while 0 indicates all the flux was removed during RDI.

We apply this throughput correction during the spectral extraction. We place oval apertures 100 au from the central star, chosen such that the PSF subtraction residual of the diffraction spikes is negligible, and the signal-to-noise is highest. These apertures have a semimajor axis of 12 pixels along the disk major axis (1.2″, equivalent to 23.5 au) and a semi-minor axis of 4 pixels (0.4″, 7.8 au) along the disk minor axis. During throughput correction, we find a throughput of 0.83 and 0.85 for the NE and SW arms, respectively (Fig. S3) and an average signal-to-noise (S/N) ~ 55 across the spectra. We perform aperture photometry across all wavelength slices using the same aperture. We find negligible effect of the convolution of the wavelength dependent PSF, so use the same size aperture for all wavelength slices. After we extract the photometry, we calculated the throughput in that aperture and divided the photometry by the throughput to correct for algorithmic flux loss. The uncertainty on the spectral extraction was determined as the standard deviation in the spectral direction (*11*), using the disk-free cube from negative disk injection. The scripts for spectral extraction are available in Data S2.

After we extracted the spectra, we removed a thermal excess at 5 μm (*39*), corresponding to ~2.8% flux decrease for $\lambda > 4.5$ μm. We also corrected for stellar color and removed photospheric absorption lines to leave only light scattered off dust particles in the disk. We did this by dividing a stellar photosphere model of β Pictoris (*57*), smoothed to the spectral resolution of the NIRSpec IFU at each wavelength, from the extracted spectra ($F_{\mathrm{star}}$) (Fig. S4). The model additionally accounted for the effects of stellar rotation and limb darkening and they were removed concurrently with the color correction and removal of the photospheric lines.

Ice map and three color image

We computed an ice map to inform our choice of spectral apertures. We synthesized three images of the disk by convolving the data cube with the passbands of three filters: a continuum image using the JWST Near Infrared Camera (NIRCam) F250M filter from 2.35 to 2.75 μm, an ice image using the Large Binocular Telescope (LBT) LBT Mid-InfraRed Interferometric Camera (LMIRCam) narrow-band $H_2O$ filter from 2.8 to 3.2 μm, and another continuum image using the LBT LMIRCam L' filter from 3.3 to 4.15 μm. The continuum filters (F250 and L') were chosen to be as close to the water ice feature without overlapping it.

To compute the ice continuum from the continuum images, we follow previous methods *(55)* to interpolate over the F250M and L' images as

$$\log F_{\mathrm{H_2O}}^{\mathrm{cont}} = \frac{\log F_{\mathrm{L'}}^{\mathrm{obs}} - \log F_{\mathrm{F250M}}^{\mathrm{obs}}}{\log \lambda_{\mathrm{L'}} - \log \lambda_{\mathrm{F250M}}} \times (\log \lambda_{\mathrm{H_2O}} - \log \lambda_{\mathrm{F250M}}) + \log F_{\mathrm{F250M}}^{\mathrm{obs}} \tag{S4}$$

where $F^{\text{cont}}_{\text{H}_2\text{O}}$ is the surface brightness at the water band continuum; $F^{\text{obs}}_{\text{F250M}}$ and $F^{\text{obs}}_{\text{L'}}$ are the observed surface brightnesses at the F250M and L' filters; and $\lambda_{\text{F250M}}$, $\lambda_{\text{H}_2\text{O}}$, and $\lambda_{\text{L'}}$ are the central wavelengths of their respective filters. Then, the optical depth at the ice feature is

$$\tau_{\text{ice}} = -\ln\frac{F^{\text{cont}}_{\text{H}_2\text{O}}}{F^{\text{obs}}_{\text{H}_2\text{O}}}, \tag{S5}$$

where $F^{\text{obs}}_{\text{H}_2\text{O}}$ is the observed surface brightness at the water band. Fig. S5 shows the wavelength combined surface brightness map over the whole 0.6 to 5 μm band. We find that the depth of the ice is approximately uniform, at $\tau_{\text{ice}} = 0.47 \pm 0.25$, across the disk. We therefore used large apertures to extract the spectra of the ice features.

To create a false color image, we mapped the three images to red, green, and blue color (RGB) channels. The blue component is F250M filter, the green component is the LMIRCam narrow-band filter, and the red component is 0.6× the L' filter. These components were then mixed proportionally across the RGB channels to create cyan (green + blue), magenta (red + blue), and yellow (red + green). We used these images to construct the false color image of the disk in Fig. 1A.

Optical depths and band profiles

To measure the optical depths and column densities of the $H_2O$ and $CO_2$ ice features, we first consider whether we can assume an optically thin disk for β Pictoris. Observations of other edge-on protoplanetary disks (*14*, *58*) have shown ice feature depths increase at the disk mid-plane, where the column density of material is highest, and decrease in the upper disk layers. Because the β Pictoris disk is edge on, we examined both the vertical and radial directions to assess the optical thickness. From the ice map (Fig. 1C), we see that the ice absorption is consistent in the radial direction, so the line of sight close to the star does not pass through substantially more material. We extract three circular apertures (radius=0.3″) along the vertical extent of the disk (Fig. S6), which show no variation in the ice profile, indicating constant optical depth. We therefore conclude that the disk is optically thin, because the column density of material is not increasing at the mid-plane.

To measure the optical depth of the $H_2O$, $CO_2$, and $^{13}CO_2$ features, we calculate

$$\tau = -\ln\frac{F_\nu}{F_{\nu,\text{cont}}}, \tag{S6}$$

where $\tau$ is the optical depth, $F_\nu$ is the observed ice feature flux, $F_{\nu,\text{cont}}$ is the flux of the local continuum. The local continuum was fitted using a third polynomial model (Fig. S7). For $H_2O$, we use the continuum in the range [2.0, 2.6] and [3.4, 4.17] μm. For $CO_2$, we use [2.0, 2.6], [3.4, 3.9], [4.32, 4.34], and [4.46, 4.6] μm, and for $^{13}CO_2$, we use [3.65, 4.17], [4.3, 4.37], and [4.46, 4.6] μm. We use laboratory band strengths of the features to convert the optical depths to column densities by

$$N = \frac{\int \tau d\nu}{\text{A}}, \tag{S7}$$

where $N$ is the column density in cm$^{-2}$, ν is the wavenumber in cm$^{-1}$, and A is the band strength. We adopt band strengths of $2.0 \times 10^{-16}$, $7.6 \times 10^{-17}$, $7.8 \times 10^{-17}$ cm molecule$^{-1}$ for $H_2O$, $^{12}CO_2$, $^{13}CO_2$, respectively (*59*).

We fitted the band profiles with a skewed Gaussian function, using the least-squares minimization lmfit package (*60*). The scripts to perform these measurements are given in Data S4.

Modeling the $H_2O$ profile

We fitted dust models to the disk reflectance spectra with varying volume fractions and compositions of water ice, following previous methods (*11*), and with scripts given in Data S3. To account for the varying effect of grain size and shape on the water ice profile, we calculate the disk scattered light at each wavelength as:

$$\frac{F_{\mathrm{disk}}}{F_{\mathrm{star}}} = \int_{a_{\min}}^{a_{\max}} g_{\mathrm{HG}}(\theta)\, Q_{\lambda,\,\mathrm{sca}}\, \pi\, a^{2+p}\, da\,, \tag{S8}$$

where $F_{\mathrm{disk}}$ is the scattered light flux, $F_{\mathrm{star}}$ is the stellar flux, $a$ is the grain size, $p$ is the index of the power-law size distribution, $g_{\mathrm{HG}}$ is the Henly-Greenstein asymmetry parameter, $\theta$ is the scattering angle, $Q_{\mathrm{sca}}$ is the scattering efficiency. Because the disk is edge-on, we assume $\theta = 90°$. We assume Mie scattering to calculate the efficiencies and asymmetry parameters of each medium, using the miepython software (*61*)

Mie theory is commonly used to model scattered light in debris disks, though it has been shown to not reproduce the total intensity and polarization intensity scattering phase functions in these disks (*62*, *63*, *64*). Mie theory assumes that dust grains are simple hard spheres, which is unlikely to reflect the grains in the β Pictoris disk.

We find $g \approx 0.9$ at 100 au. This is similar to previous measurements of $g \approx 0.8$ to $0.9$ in the outer disk of β Pictoris (*52*), indicating strong forward scattering. Other models (*53, 54*) found lower values of $g$ or required a double HG function to model the inner disk of β Pictoris and has been interpreted as an indication of a gradient in the dust grain properties (*52*).

We consider multiple dust species mixed together to form a homogeneous effective medium and calculate the optical constants assuming Bruggeman's rule (*65*):

$$\Sigma_{S=1}^{S} V_S \frac{n_S^2 - n_{\mathrm{eff}}^2}{n_S^2 + 2n_{\mathrm{eff}}^2} \tag{S9}$$

with

$$\Sigma_{S=1}^{S} V_S = 1. \tag{S10}$$

$S$ is the total number of dust components, $s$ is the index of each dust component, $V_S$ is the component volume fraction, $n_{\mathrm{eff}}$ is the effective optical constant of each component, and $n_s$ is the optical constant of each component. We assume a porosity as a dust component with the optical constants of a vacuum ($n_s$= 1).

Because icy grains are composed of ice covering silicate or hydrogenated amorphous carbon grains, we utilize laboratory optical constants of amorphous olivine (MgFeSiO4) (*66*) and amorphous carbon (*67*). Amorphous olivine is consistent with silicate features observed in β Pictoris in the mid-infrared (*57*). Spectral energy distribution (SED) models have shown that a mixture of 60% silicates to 40% amorphous carbon in non-porous grains can reproduce the dust in the disk of β Pictoris (*68*). For the ice optical constants, we looked at three different types of water ice: amorphous water ice at 50 K, crystalline water ice at 50 K, both of which are consistent with the dust temperature at 100 au, and a 10:1 mixture of water and $CO_2$ ice at 40 K, which is based on our $^{13}CO_2$ spectral fitting and abundance measurements. We computed optical constants for the pure water ice (*29*) and $CO_2$ (*69*) using the optool software (*70*). The optical constants for the water and $CO_2$ mixture (*71*) were derived from the transmission spectra using the icenk software (*72).*

Using these optical constants, we fitted the values of $a_{\min}, p, V_S$, and the proportion of the dust populations for $S > 1$. We set the porosity $V_P = 0$, maximum size $a_{\max} = 1$ mm, and carbon volume fraction to be 0.66 of the olivine volume fraction, $V_C = 0.66V_{\text{oli}}$, and explore the ranges [0, 5] μm for $a_{\min}$, [-4, -2] for $p$, and [0, 1] for each $V_S$. The proportion of the last component is set by $1 - \Sigma_1^{S-1} V_S$. We use differential evolution (DE) optimization with an initial population number, $N_{\text{pop}} = 10 \times$ the number of free parameters $M_{\text{p}}$, to fit the model to the spectral $H_2O$ data (Fig. 3). The disk flux is calculated from 2.2 to 3.5 μm, to span the water ice band. The disk flux is calculated and compared to the spectral extraction by computing

$$\chi_r^2 = \frac{\chi^2}{\nu} = \frac{1}{\nu} \Sigma_S \frac{(F_d - F_m)^2}{\sigma_{\text{d}}^2}, \tag{S11}$$

where $\chi_r$ is the reduced chi-square, $\chi$ is the chi-square, $\nu = N_{\text{wave}} - M_{\text{p}}$ is the degrees of freedom, $N_{\text{wave}}$ is the number of wavelength channels, $F_d$ is the flux of the data, $F_m$ is the flux of the model, $\sigma_d$ is the uncertainty of the data. We mutate the population following the "best/1/bin" strategy, where 'best' indicates the base vector used is the best vector in the population at the point in the algorithm, '1' refers to using only one vector difference to form the mutated population, and 'bin' refers to applying a uniform crossover when forming the trial population (*55*). We use a mutation constant between 0.5 to 1.0 until the DE algorithm minimizes $\chi_r^2$ and converges on a best-fitting solution. We then calculate the optical depth of the model following the method outlined above (using Equation S6).

Table S1 lists the results for six different models shown in Fig. 3, and Figs. S8-S13 show the DE optimization corner plots. We consider models to be consistent to the observations if $\chi_r^2 \leq \chi_{r,\min}^2 + \sqrt{2/\nu}$ (*13*, *56*), where $\chi_{r,\min}$ is the minimum reduced chi-square. There is some degeneracy in determining the dust compositions (*11*), due to the limited spectral coverage, low spectral resolution, and lack of observed silicate spectral features.

Crystallinity fraction

To estimate the crystallinity fraction using the 3.1 μm Fresnel peak, we produced models of AWI+CWI+$CO_2$ using the average dust parameters of our best fitting water ice models (AWI+$CO_2$ and AWI+CWI+$CO_2$) while varying the AWI and CWI volume fractions. We use the best fit parameters found from our AWI+CWI+CO2 model given in Table S1: $a_{\min} = 2.33$ μm, $p = -3.76$, $V_{\text{oli}} = 0.099$, $V_{\text{C}} = 0.066$, and $CO_2$ volume fraction $V_{CO_2} = 0.2214$. These volume fractions implies the sum of the AWI volume fraction ($V_{\text{AWI}}$) and CWI volume fraction ($V_{\text{CWI}}$) $V_{\text{AWI}} + V_{\text{CWI}} = 0.613$. We vary the amorphous water ice volume fraction from 0.0 to 0.61, in increments of 0.01, while modeling the ice profile as in the previous section. We then convert the profile to optical depth and remove the local continuum around the Fresnel peak to normalize the spectra. We calculate the crystallinity fraction

$$f_{\text{cry}} = \frac{V_{\text{CWI}}}{V_{\text{CWI}} + V_{\text{AWI}}}, \tag{S12}$$

where $V_{\text{CWI}} = 1 - V_{\text{oli}} - V_{\text{C}} - V_{CO_2} - V_{\text{AWI}}$. Fig. S14 shows the modeled Fresnel peaks for varying values of $f_{\text{cry}}$. We then calculated the normalized offset between the location of the Fresnel peak in the model and the average observed value between 3.11 to 3.125 μm (the range where the modeled peaks are located). We found that the Fresnel feature would be detected at three times the uncertainty in the NIRSpec data (Fig. S14B), for a crystallinity factor of 0.068 (equivalent to $V_{\mathrm{CWI}} = 0.042$).

Modeling the $^{13}CO_2$ feature

Previous observations of the $^{12}CO_2$ ice feature at 15.2 μm (*36*) and the $^{13}CO_2$ band at 4.39 μm (*28*) have decomposed those features into up to five components, each due to $CO_2$ in a different ice matrix. The $^{13}CO_2$ feature is not affected by grain size effects, so we compare laboratory spectra to the observations without accounting for distortions due to size or grain shape. However, our observations have a wavelength dependent resolving power of ~200 at 4.39 μm and S/N ~ 7 to 11, which is not sufficient to decompose the band into multiple components. We therefore seek to determine only the dominant ice component.

The peak position of 4.389 μm and the lack of CO ice features indicate that the dominant ice component is likely above 30 K (as CO ice freezes at this temperature). We do not detect other ice or organic features in the spectra, such as $CH_3OH$ ice at 3.53 μm. Therefore, we consider only mixtures of $H_2O$, $CO_2$ and CO ices. We focus on laboratory mixtures with astrophysically plausible abundance ratios, using Solar System minor bodies as an empirical reference (*25*).

Following previous work (*28*), we compare our observations to spectra of various $CO_2$ ice mixtures in the Leiden Ice Data Base for Astrochemistry (*73*). We write these mixtures as relative ratios of the molecules in the mixtures, e.g. an equal 1 to 1 mixture of $CO_2$ to CO is written as CO:$CO_2$=1:1. We use pure $CO_2$ from 15-80 K (*70*), CO:$CO_2$=1:1 at 15 to 90 K (70), CO:$CO_2$=2:1 at 15 to 80 K (*74*), $CO_2$:$H_2O$=1:10 at 10 to 170 (*71*), $H_2O$:$CO_2$:CO=100:20:23 at 20 K (*71*), and $CO_2$:$CH_3OH$=3:1 at 10 to 112 K (*71*). In each spectrum, the $^{13}CO_2$ band overlaps with the tail of the $^{12}CO_2$ band, introducing excess optical depth. We therefore fitted the tail of the $^{12}CO_2$ band in the ranges 2250 to 2260 cm$^{-1}$ and 2290 to 2296 cm$^{-1}$ using a linear model, then subtracted it from the spectra. We then convolved and smoothed the lab spectra to the NIRSpec IFU wavelength dependent resolving power, which at 4.39 μm is $R$ = 217, and normalized the spectra to the peak of the $^{13}CO_2$ feature (Fig. 4).

For the observed spectrum from the NE aperture, we determined the optical depth of the $^{13}CO_2$ feature as described above, then normalized it to the peak of the feature. We fitted a Gaussian model to the observed $^{13}CO_2$ feature to determine the width of the likely major component; the peak position was found from the optical depth fitting discussed above. For $CO_2$:$H_2O$, we determined the temperature that best matches the observed peak position by determining the temperatures where the model peak overlapped the observed peak. For $CO_2$:$H_2O$=1:10, this ranged from 10 to 50 K, with the best fit at 40 K. For CO:$CO_2$=1:1, the best fitting temperature is 30 K, though this does not match the peak, likely excluding the CO:$CO_2$ mixture as the dominant ice component. Although the $H_2O$:$CO_2$:CO=100:20:23 mixture can match the peak position, the lack of CO ice in the spectra indicates this is likely not the dominant component.

To compare our best fit temperatures to the blackbody grain temperature, we determine the dust temperature by assuming radiative equilibrium (*39*). In radiative equilibrium, the energy absorbed by dust grains (as a function of distance from the star) is equal to the energy emitted by a black body at the dust temperature:

$$\int_0^\infty Q_{\mathrm{abs},\lambda} J_\lambda(T_*)\mathrm{d}\lambda = \int_0^\infty Q_{\mathrm{abs}}(\lambda)\, B_\lambda(T_\mathrm{d})\mathrm{d}\lambda, \tag{S13}$$

where $Q_{\mathrm{abs},\,\lambda}$ is the absorption coefficient as a function of wavelength for the ice material, $J_\lambda(T_*)$ is the stellar mean intensity as a function of stellar temperature ($T_*$) and wavelength, $B_\lambda(T_\mathrm{d})$ is the black body Planck function for grain emission with temperature $T_\mathrm{d}$. In a blackbody limit $Q_\mathrm{abs} \sim 1$, a solution to equation S13 is

$$T_d = \left(\frac{R_*^2}{4D^2}\right)^{1/4} T_*, \tag{S14}$$

where $R_*$ is the stellar radius of 1.65 stellar radii ($R_\odot$) for β Pictoris (*75*), $D$ is the distance from the star, and $T_*$ is the stellar temperature of 8000 K (*75*). For distances of 90 to 110 au, we find temperatures of 45 to 50 K.

The estimated temperature of $H_2O$ dominated $^{13}CO_2$ ice is calculated based on the peak position as a function of temperature in laboratory spectra. The $^{13}CO_2$ peak position depends on temperature for different mixtures of $H_2O$:$CO_2$ and follows the relation (*40*)

$$\frac{\mu}{\mathrm{cm}^{-1}} = (2280.16 \pm 0.06) - (0.03 \pm 0.0006)T, \tag{S15}$$

where $\mu$ is the peak position and $T$ is the laboratory ice temperature. Using our measured peak positions of 2278.3± 0.58 $\mathrm{cm}^{-1}$ and 2278.8 ± 1.13 $\mathrm{cm}^{-1}$ for the NE and SW arms, respectively, this matches the laboratory trend at ~50 ± 8 K (*40*). As the timescales between the laboratory and astrophysical temperature measurements greatly differ, we use a conversion factor between laboratory and astrophysical temperatures of 0.4 ± 0.1 to convert the ice temperature in the laboratory timescale to an astrophysical temperature of 21 ± 8 K (*40*).

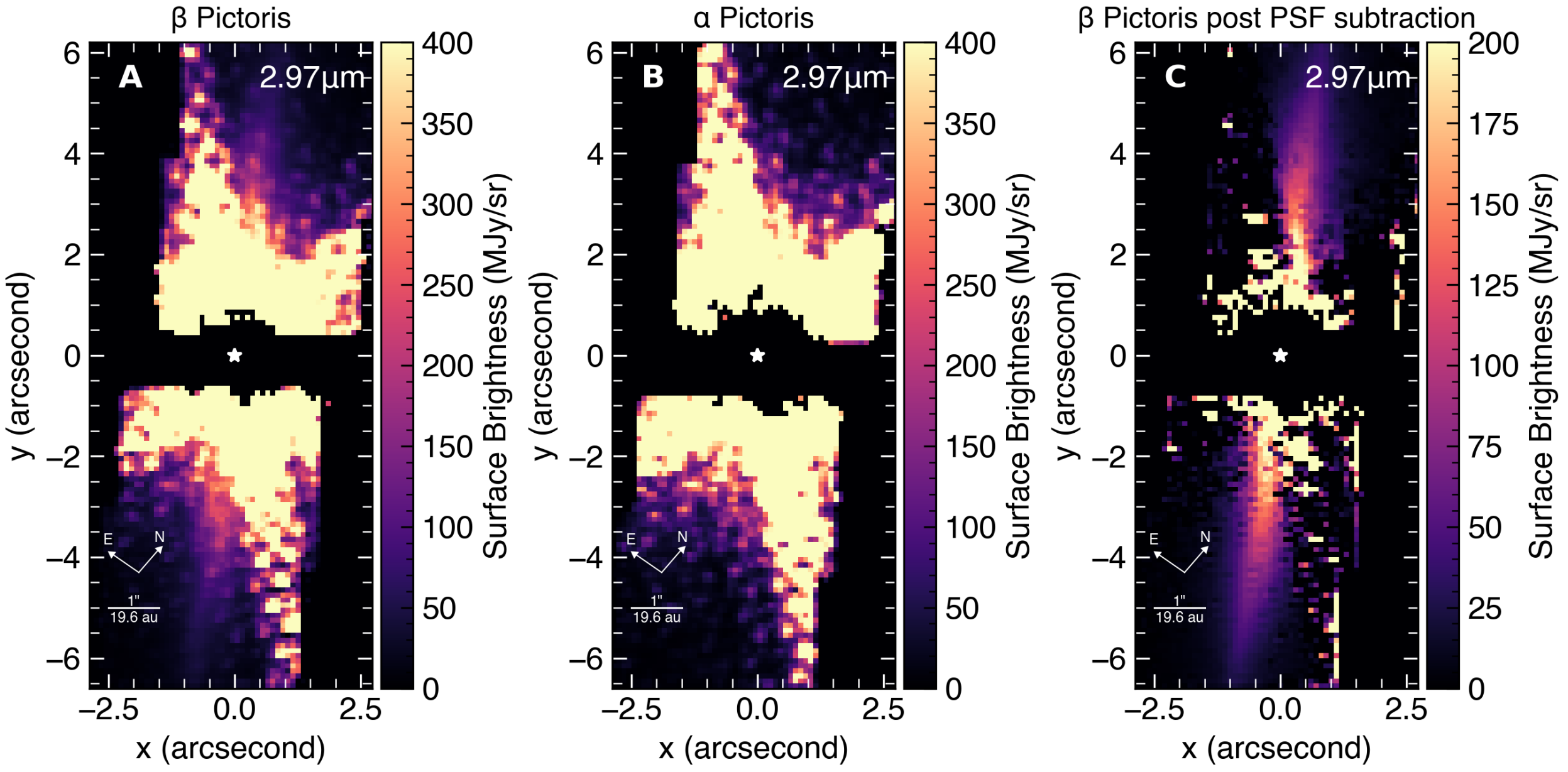

**Fig. S1. PSF subtraction process.** **(A)** The 2.97 μm slice of the β Pictoris IFU cube before PSF subtraction. **(B)** Same as A but for the PSF reference star, α Pictoris. **(C)** The 2.97 μm slice of the β Pictoris cube after PSF subtraction.

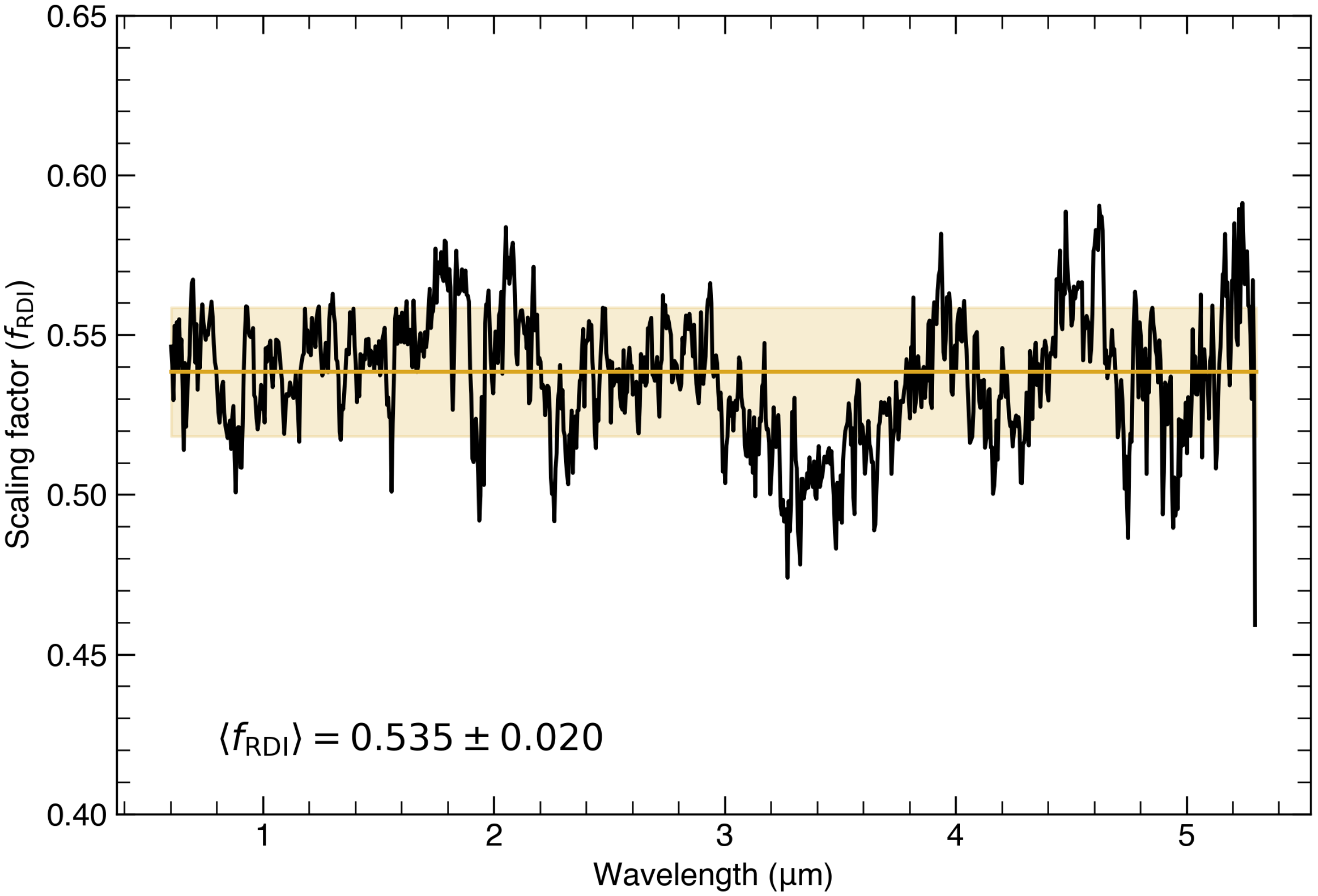

**Fig. S2 RDI scaling factor as a function of wavelength.** The black line shows the scaling factor for RDI as a function of wavelength. We find it is stable within 2%. We use the average value (gold line) to perform PSF subtraction. The light gold shading is the 1σ uncertainty on the average value. Using a wavelength dependent scaling factor introduces high frequency noise into the spectrum. We do not see any correlation between the wavelength dependent scaling factor and the spectral ice features in the extracted disk spectra.

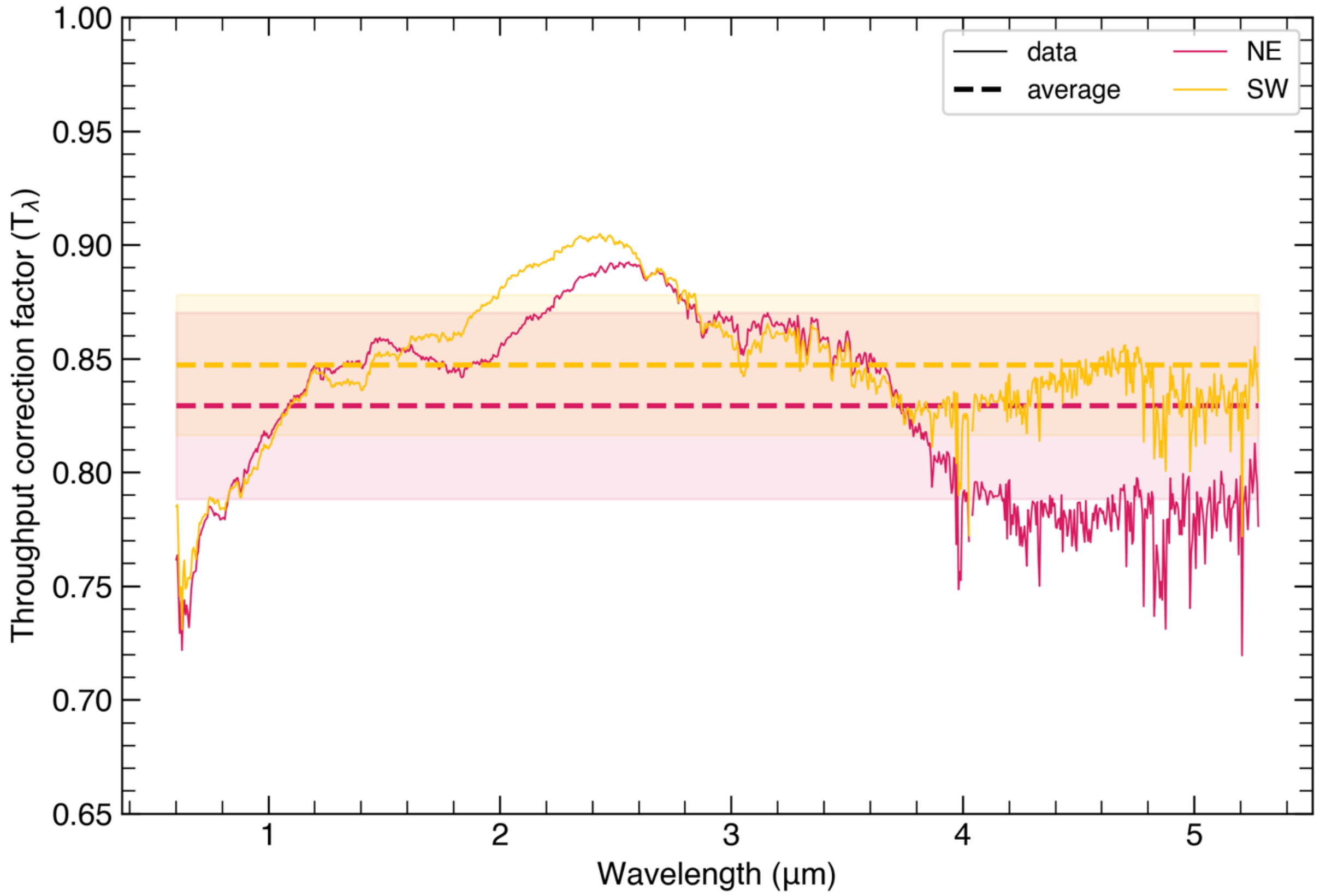

**Fig. S3. RDI throughput correction factor.** The throughput factors for the NE (magenta solid line) and SW (gold solid line) aperture extractions are plotted as a function of wavelength. The dashed lines are the average of the data (solid lines) with the 1σ uncertainty shown as the shaded regions. The flux loss is less than 10% across wavelengths. Using a wavelength dependent throughput introduces high frequency noise into the spectrum.

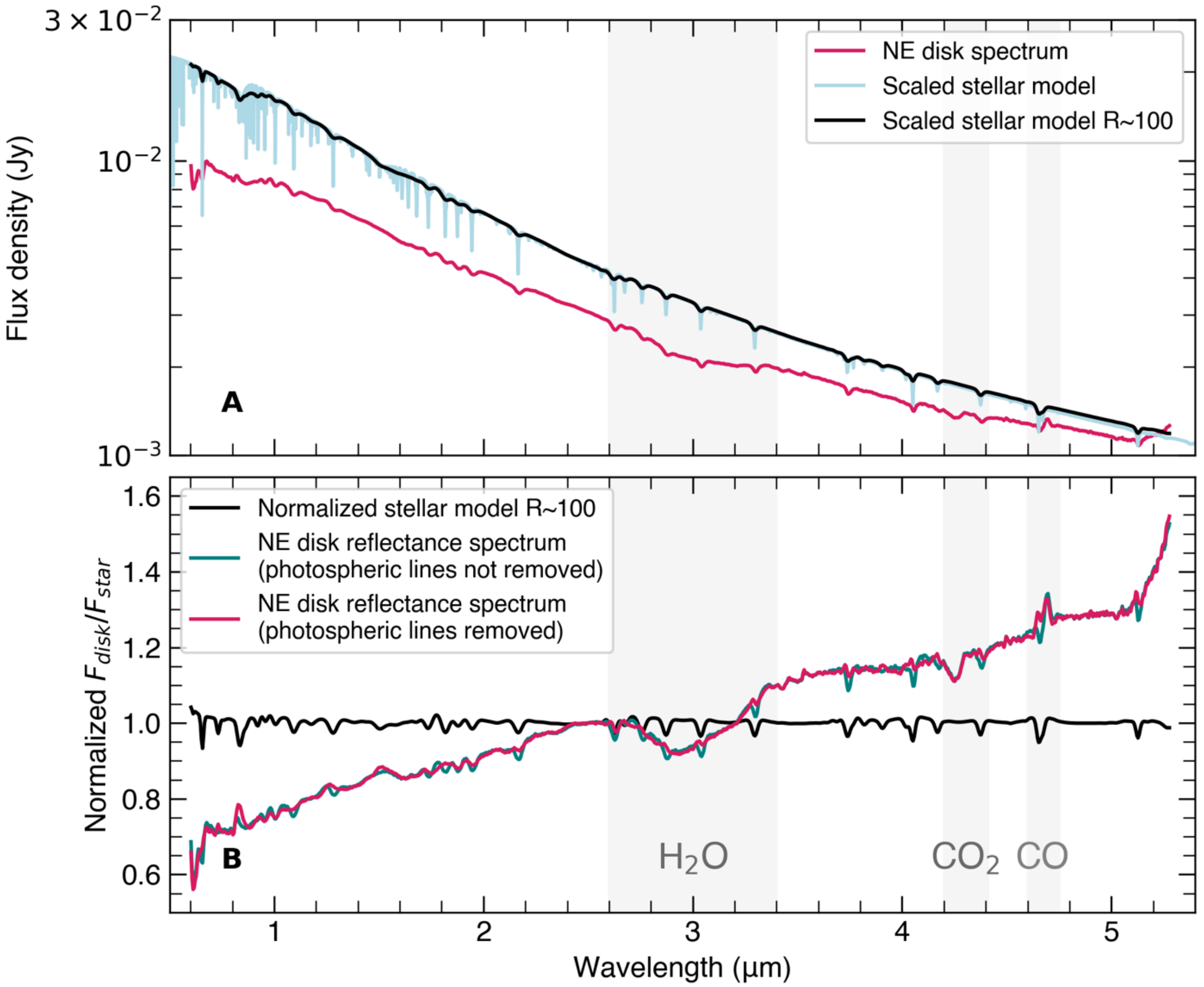

**Fig. S4. Example of removal of stellar light from the disk spectra.** In both panels, gray shaded bands highlight the locations of the $H_2O$, $CO_2$, and CO features. **(A)** Comparison between the NE aperture extracted disk spectrum after RDI (magenta) and a scaled stellar model of β Pictoris at its original resolution (blue) and smoothed to R ~ 100 (black). The y-axis flux density is in units of Jansky (Jy). The photospheric absorption lines are visible. **(B)** The disk reflectance spectrum after subtraction of the R ~ 100 stellar model and normalized at 2.5 μm (magenta) compared to the normalized smoothed stellar model (black). For comparison, the cyan line shows the disk reflectance spectra without the photospheric absorption lines removed.

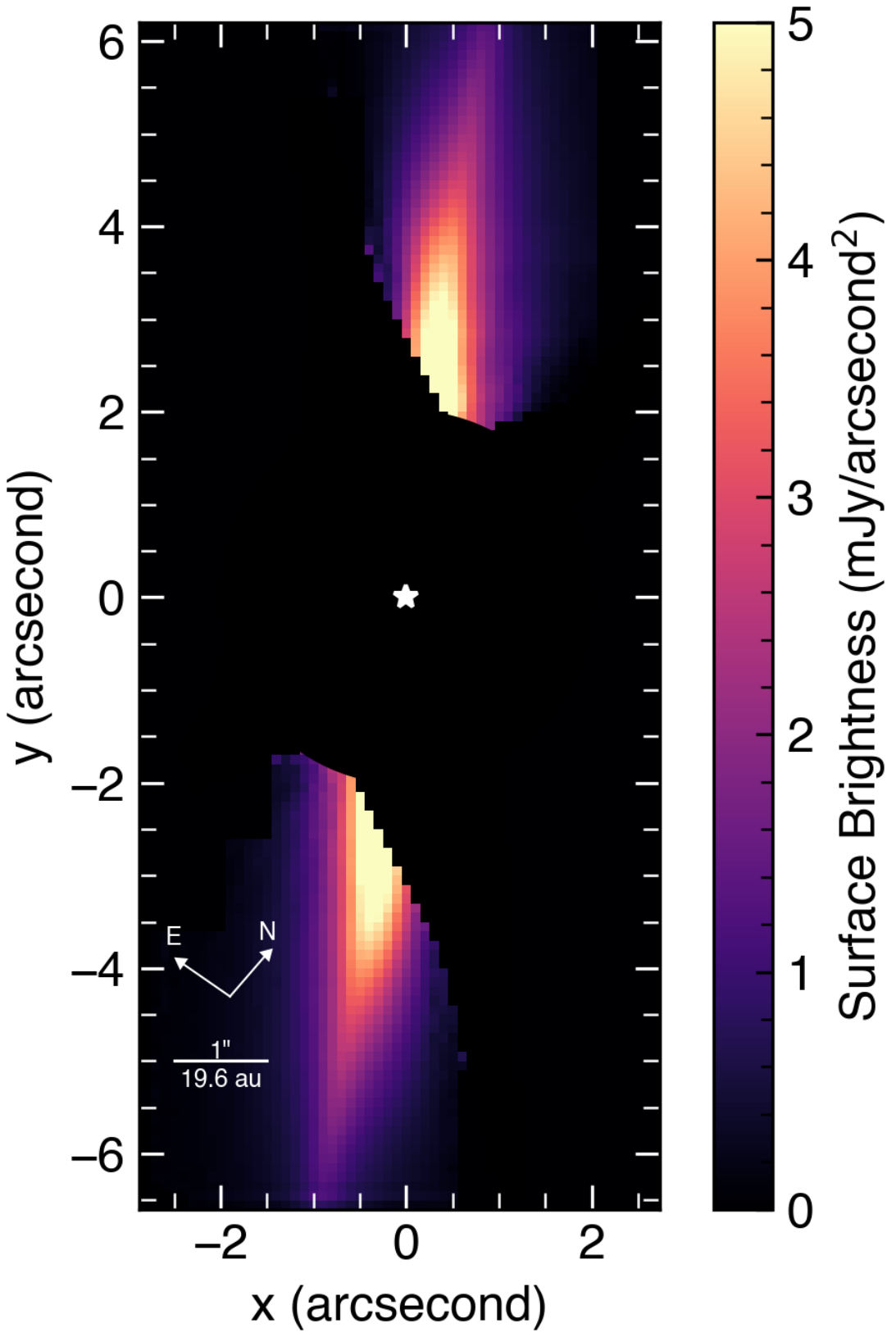

**Fig. S5. Surface brightness map of β Pictoris.** Same as Fig. 1A but showing the surface brightness of the β Pictoris disk, measured from 0.6 to 5.2 μm after the removal of the stellar PSF.

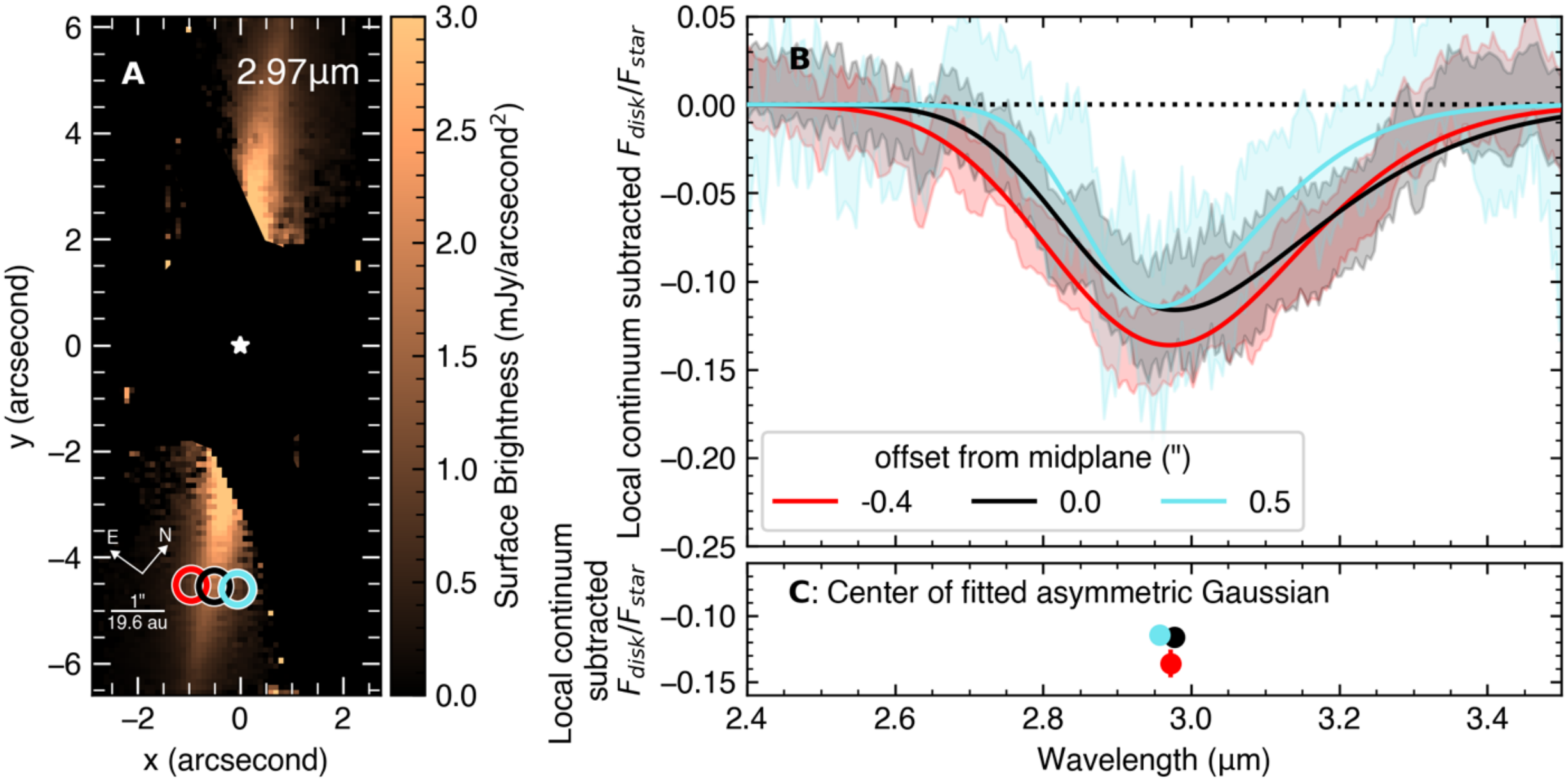

**Fig. S6. Ice depth as a function of vertical height in the disk. (A)** Same as Fig. S5 but for a 2.97 µm wavelength slice of the NIRSpec IFU data cube. Red, black, and light blue circles show apertures located at different vertical heights within the disk. **(B)** The local continuum-subtracted $H_2O$ profile from the three apertures shown in panel A (shading spanning the $\pm1\sigma$ uncertainty, colors match panel A). Colored lines are skewed Gaussian profiles fitted to the data. The black dotted line is a horizontal reference line at 0. **(C)** Data points are the peak wavelengths of the fitted skewed Gaussians in panel B with error bars representing the 1σ uncertainty in the disk flux, in matching colors.

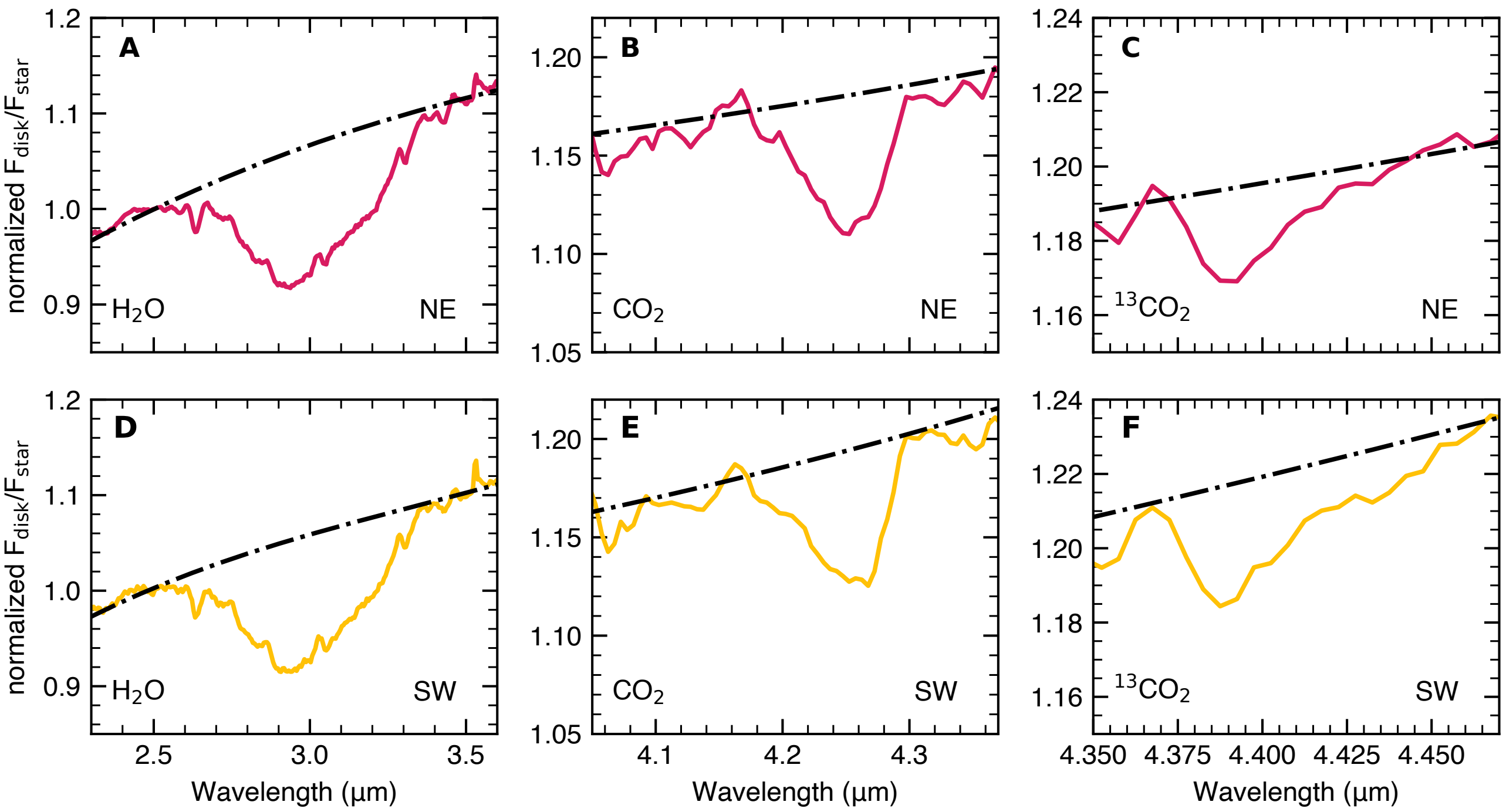

**Fig. S7. Local continua.** The local continuum (black dash-dot line) fitted to the **(A)** $H_2O$, **(B)** $CO_2$, and **(C)** $^{13}CO_2$ ice features in the NE aperture (magenta lines). **(D-F)** same as panels A-C but for the SW aperture (gold lines).

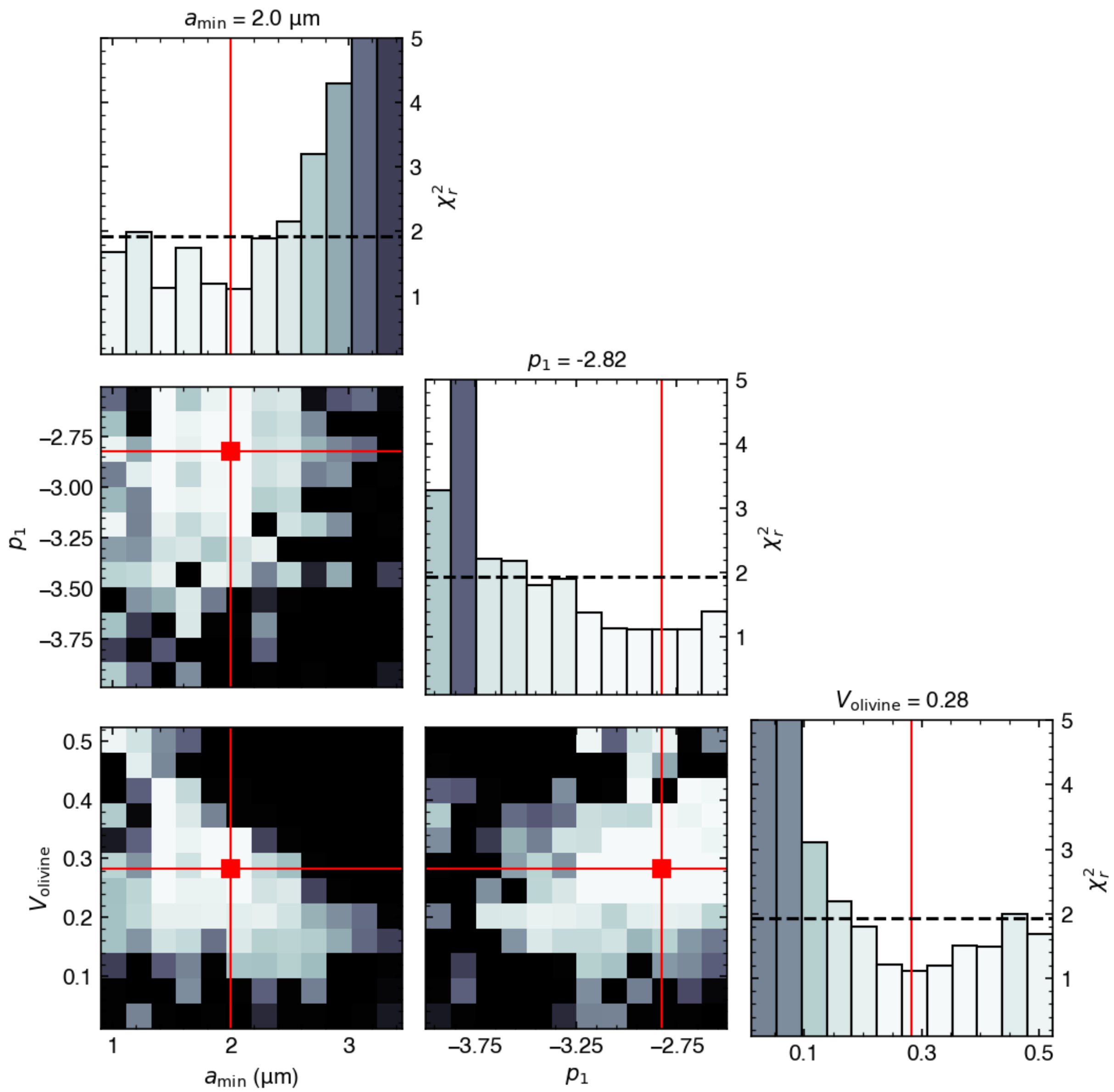

**Fig. S8. Corner plot of dust parameters fitted in the DE optimization, for the AWI model**. The off-diagonal panels show $\chi_r^2$ in grayscale for each value of each parameter, where the lighter color indicates the location where the model converges. The red dot, red lines, and value above each column indicate the best fitting value for the parameter. The diagonal panels show histograms of $\chi_r^2$ for each parameter. Lighter gray indicates lower $\chi_r^2$, and therefore the best fit solution. We regard solutions as acceptable if $\chi_r^2 \leq \chi_{r,min}^2 + \sqrt{2/\nu}$ (below the dashed black line, see Table S1).

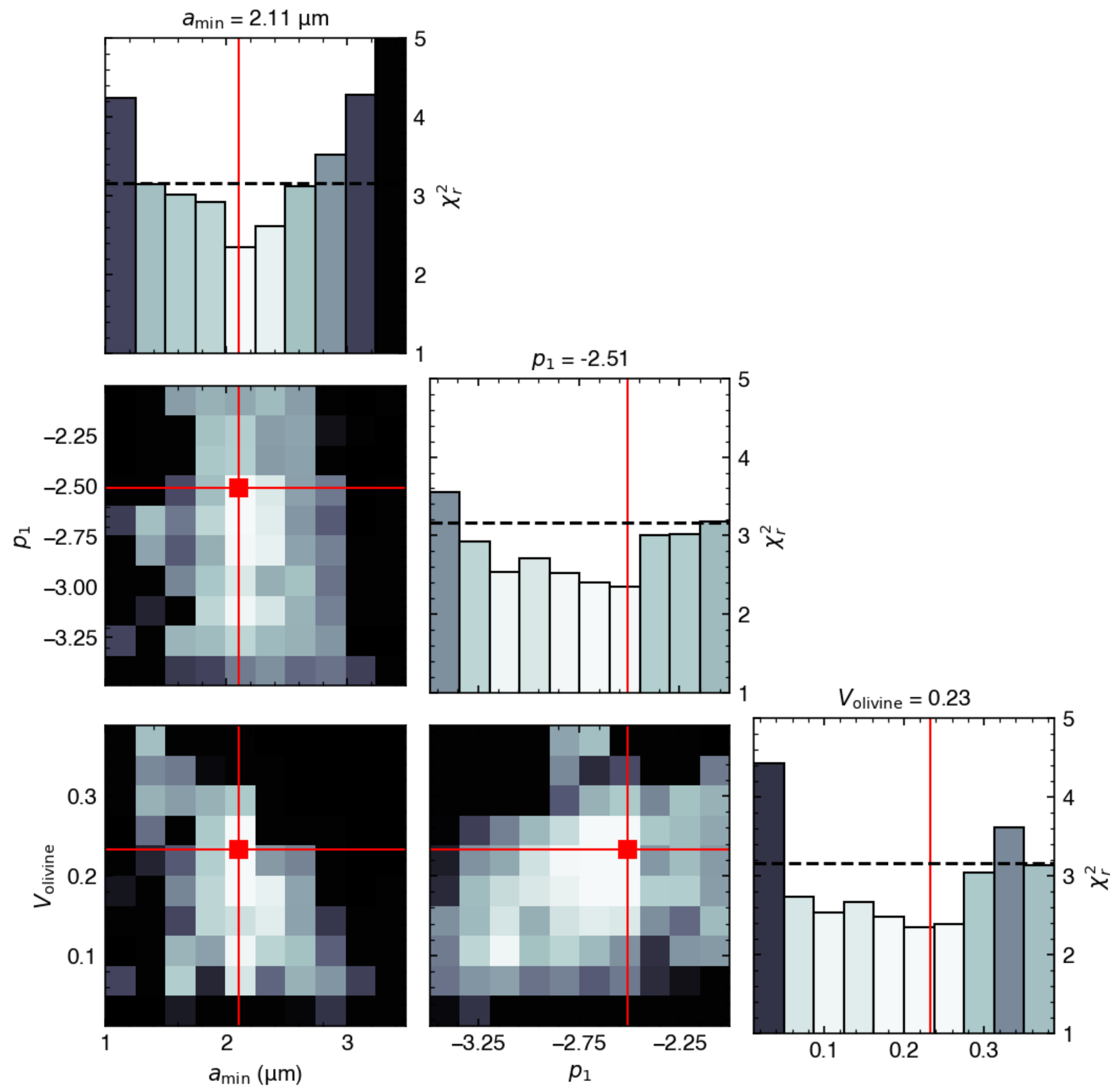

**Fig. S9. Same as Fig. S8 but for the CWI model**.

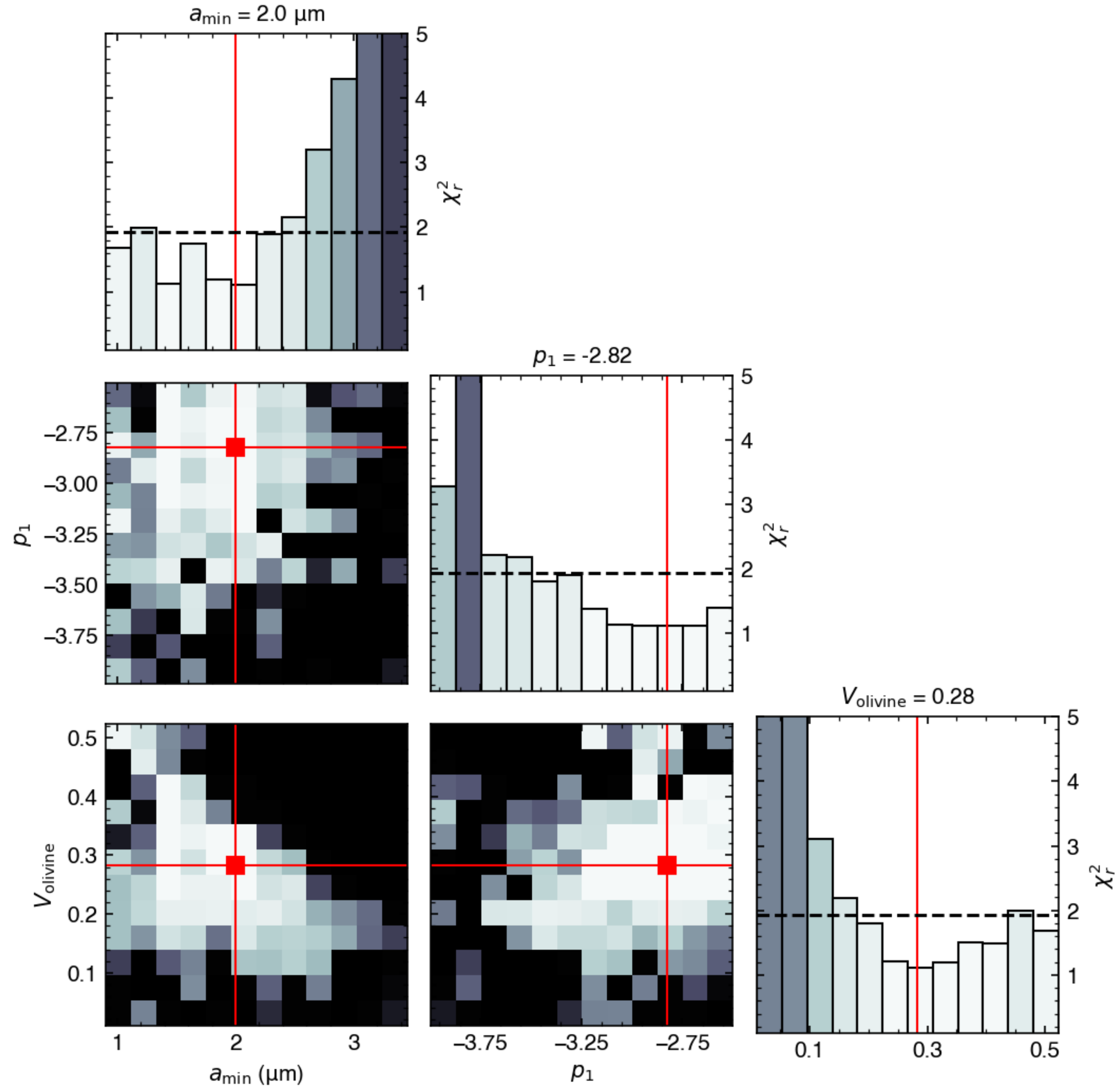

**Fig. S10. Same as Fig. S8 but for the AWI+$CO_2$ model**.

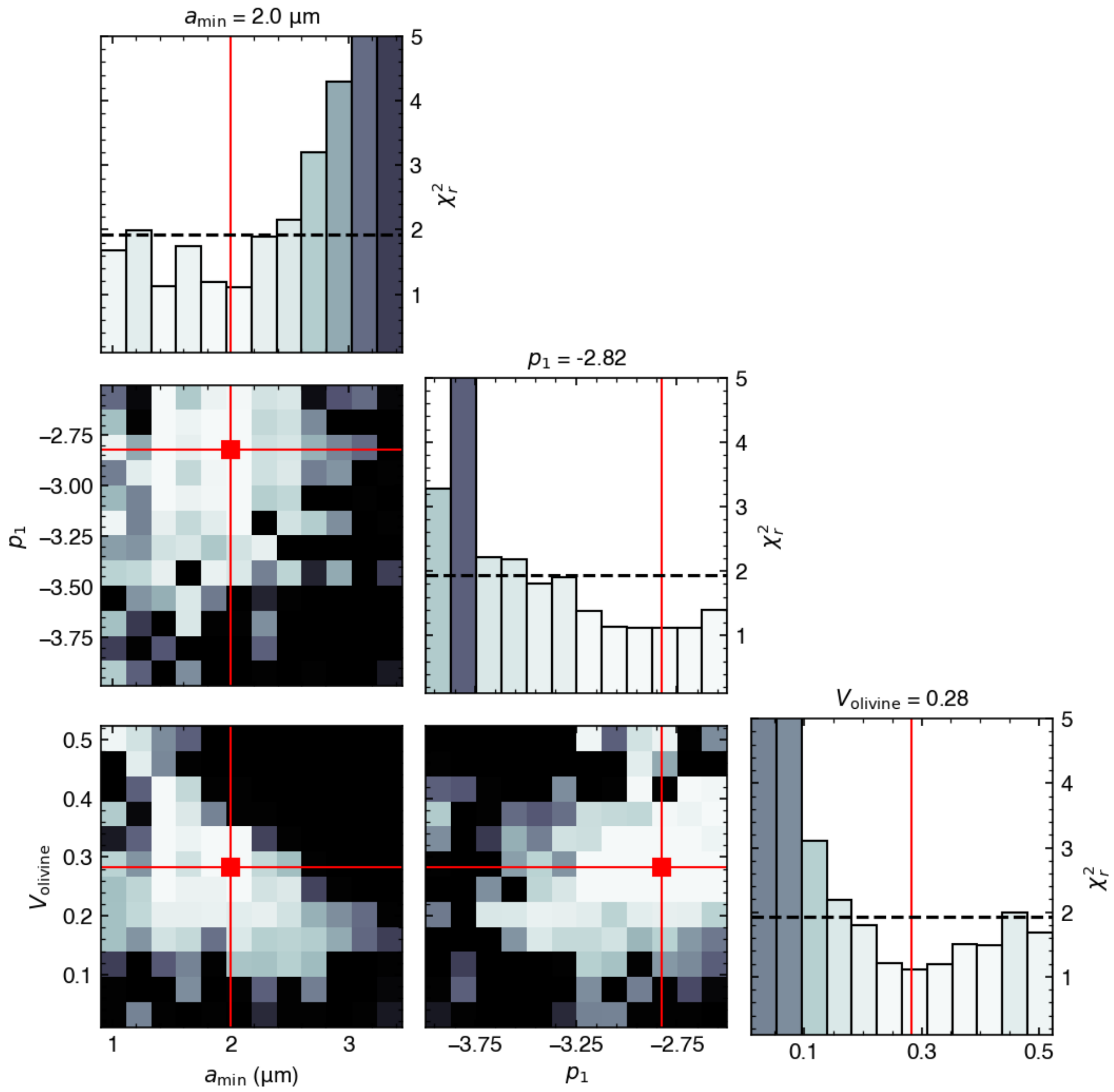

**Fig. S11. Same as Fig. S8 but for the CWI+$CO_2$ model**.

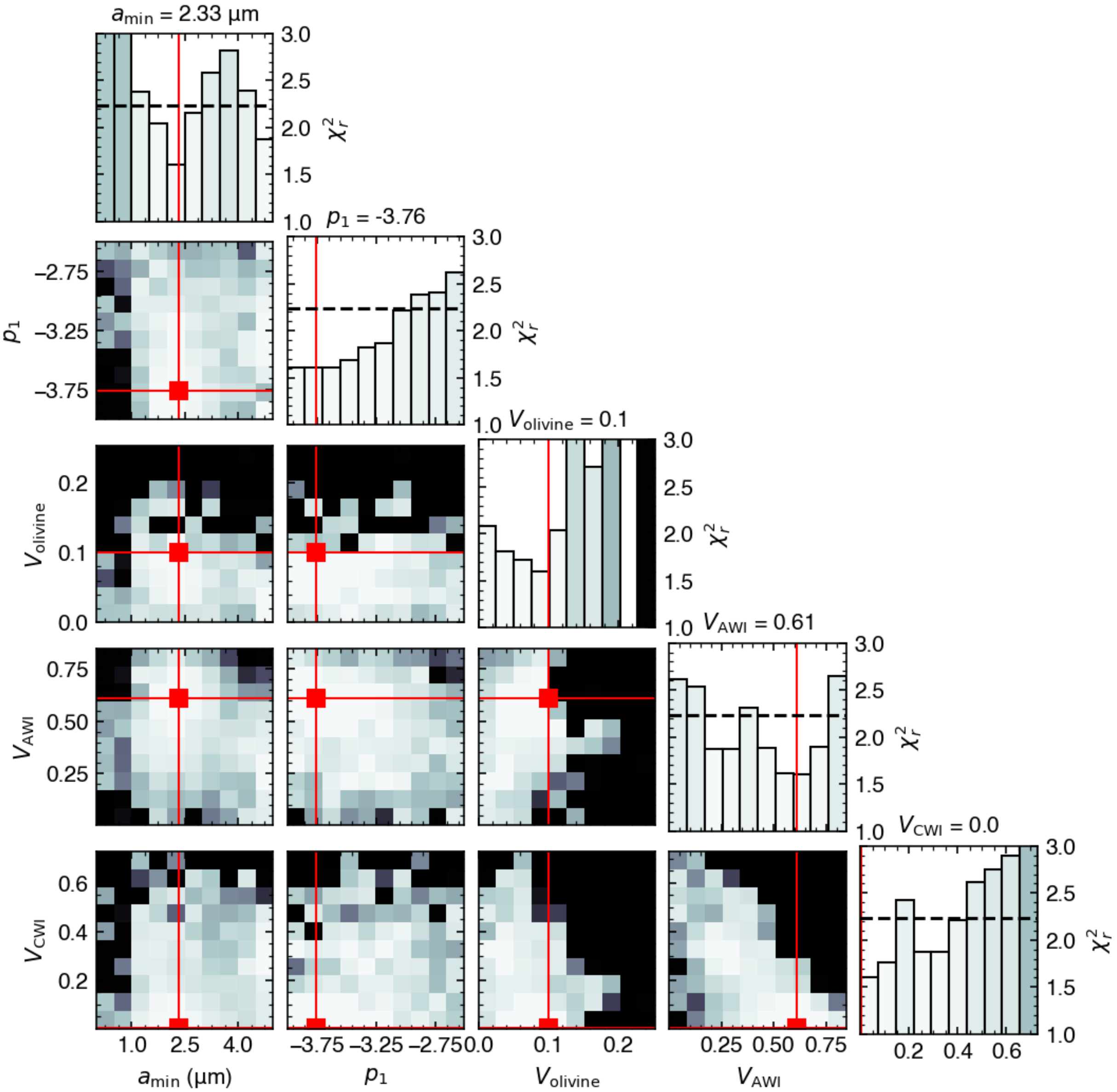

**Fig. S12. Same as Fig. S8 but for the AWI+CWI+$CO_2$ model**.

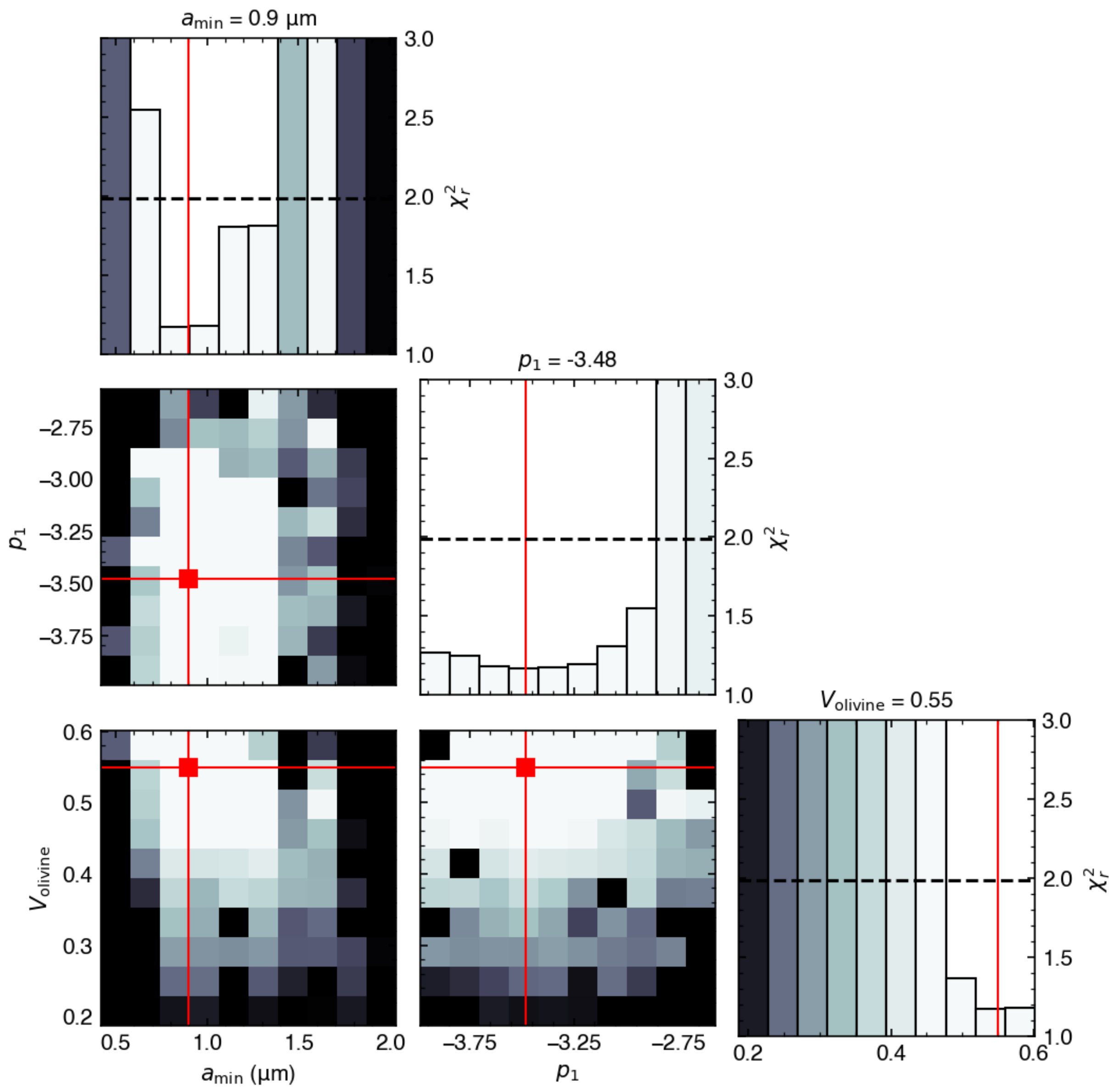

**Fig. S13. Same as Fig. S8 but for the $H_2O$:$CO_2$=10:1 model**.

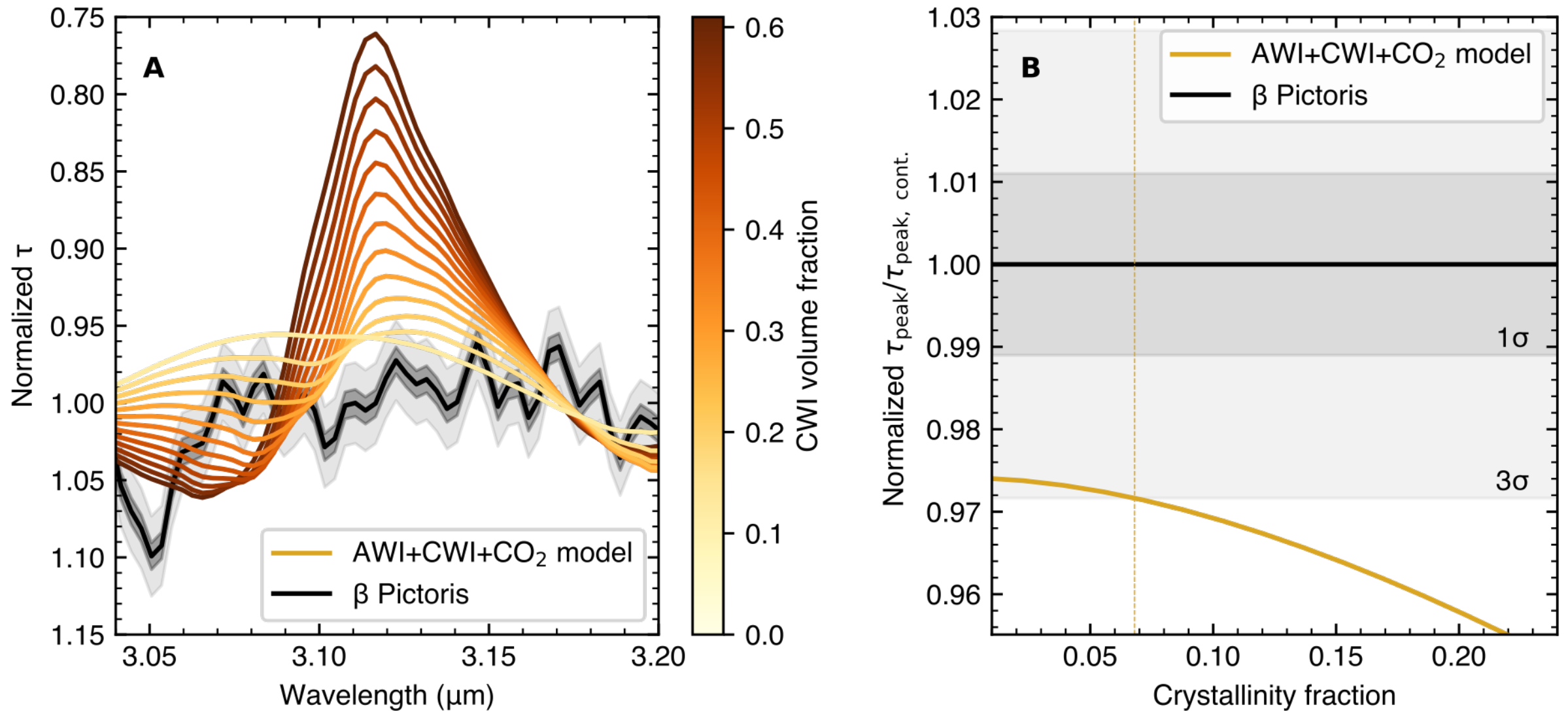

**Fig. S14. Crystallinity fraction**. **(A)** The normalized optical depth of the observations in the NE aperture (black) around the expected position of the Fresnel peak. The dark and light gray shading indicate 1σ and 3σ uncertainty, respectively. Colored lines are models of aggregate olivine, amorphous carbon, AWI, CWI and $CO_2$ ice for varying volume fractions of crystalline water ice (color bar). **(B)** The offset between the model (gold line) and the observed normalized optical depth (black line with dark and light gray shading indicating the 1σ and 3σ uncertainty, respectively from panel A), evaluated at the model peak. The intersection between the 3σ uncertainty region and the gold line determines the upper limit on the crystalline fraction (gold dashed line).

**Table S1. Best fit $H_2O$ model grain composition by volume fraction for each model.** The reduced chi-square ($\chi_r$), minimum grain size ($a_{min}$), grain size distribution exponent ($a_p$), and refractory and ice volume fractions for each model. The top row is the best fit value, while the bottom row is the range of acceptable solutions, those with $\chi_r^2 \leq \chi_{r,min}^2 + \sqrt{2/\nu}$.

| | | | | Refractory Components | | Ice Components | | |
|---|---|---|---|---|---|---|---|---|
| Model | $\chi_r^2$ | $a_{min}$ (μm) | $p$ | $MgFeSiO_4$ | C | CWI | AWI | $CO_2$ |
| AWI | 1.12 | 2.0013<br>0.97 to 2.16 | -2.82<br>-3.44 to -2.54 | 0.28<br>0.22 to 0.51 | 0.19 | - | 0.53 | - |
| CWI | 2.34 | 2.11<br>1.68 to 2.36 | -2.51<br>-3.68 to -2.15 | 0.23<br>0.073 to 0.28 | 0.16 | 0.61 | - | - |
| AWI+$CO_2$ | 1.12 | 1.29<br>1.25 to 2.35 | -2.44<br>-3.33 to -2.087 | 0.078<br>0.028 to 0.099 | 0.052 | - | 0.54<br>0.45 to 0.67 | 0.33 |
| CWI+$CO_2$ | 2.25 | 2.071<br>1.15 to 2.50 | -3.47<br>-3.94 to -3.024 | 0.078<br>0.0065 to 0.099 | 0.052 | 0.802<br>0.69 to 0.87 | - | 0.067 |
| AWI+CWI+$CO_2$ | 1.60 | 2.33<br>1.69 to 4.96 | -3.76<br>-3.99 to -2.98 | 0.099<br>0.19 to 0.70 | 0.066 | 0.0036<br>0.0 to 0.38 | 0.61<br>0.19 to 0.70 | 0.2214 |
| $H_2O:CO_2$=10:1 | 1.17 | 0.90<br>0.86 to 1.26 | -3.48<br>-3.99 to -2.98 | 0.55<br>0.50 to 0.59 | 0.36 | 0.090 | | - |

**Caption for Data S1. (.py file)**
Data analysis python script developed for Reference Differential Imaging with PSF subtraction for the JWST/NIRSpec PRISM IFU observations of β Pictoris. The script ingests calibrated and recentered data cubes, subtracts the stellar PSF, and performs disk modeling for throughput correction.

**Caption for Data S2. (.py file)**
Spectral extraction python script developed to extract aperture photometry. The script ingests the PSF subtracted data cubes, the best fitting models, and the uncertainty derived by Data S1 and extracts the spectrum over the data cube, in each defined aperture.

**Caption for Data S3. (.py file)**
$H_2O$ spectral model fitting python script developed to model the water ice optical depth and calculate the crystallinity fraction. The script ingests an optical depth spectrum and performs a differential evolution fitting using Mie theory with water ice, olivine, amorphous carbon, and carbon dioxide optical constants.

**Caption for Data S4. (.py file)**
Optical depth script developed to calculate the optical depth, peak position, and effective column density of the ice features.